\documentclass[reprint, twocolumn, amsmath, amssymb, aps, prx]{revtex4-2}

\usepackage{amsmath}
\usepackage{upgreek}
\usepackage{xcolor}

\usepackage{graphicx}
\usepackage{dcolumn}
\usepackage{bm}

\begin{document}

\preprint{APS/123-QED}

\title{\textbf{Microscopic contact line dynamics dictate the emergent behaviors of particle rafts} 
}%

\author{Ranit Mukherjee} 
\author{Zih-Yin Chen}%
\affiliation{Department of Mechanical Engineering, University of Minnesota, Minneapolis 55455, Minnesota, USA}%

\author{Xiang Cheng}%
\affiliation{Department of Chemical Engineering and Materials Science, University of Minnesota, Minneapolis 55455, Minnesota, USA}%
\affiliation{
St. Anthony Falls Laboratory, University of Minnesota, Minneapolis 55414, Minnesota, USA
}%

\author{Sungyon Lee}%
 \email{Contact author: sungyon@umn.edu}
\affiliation{Department of Mechanical Engineering, University of Minnesota, Minneapolis 55455, Minnesota, USA}%
\affiliation{
St. Anthony Falls Laboratory, University of Minnesota, Minneapolis 55414, Minnesota, USA
}%

\date{\today}

\begin{abstract}
Fluid-fluid interfaces laden with discrete particles behave curiously like continuous elastic sheets, leading to their applications in emulsion and foam stabilization. Although existing continuum models can qualitatively capture the elastic buckling of these particle-laden interfaces---often referred to as particle rafts---under compression, they fail to link their macroscopic collective properties to the microscopic behaviors of individual particles. Thus, phenomena such as particle expulsion from the compressed rafts remain unexplained. Here, by combining systematic experiments with first-principle modeling, we reveal how the macroscopic mechanical properties of particle rafts emerge from particle-scale interactions. We construct a phase diagram that delineates the conditions under which a particle raft collapses via collective folding versus single-particle expulsion. Guided by this theoretical framework, we demonstrate control over the raft's failure mode by tuning the physicochemical properties of individual particles. Our study highlights the previously overlooked dual nature of particle rafts and exemplifies how collective dynamics can arise from discrete components with simple interactions. 
\end{abstract}

\maketitle


\section{\label{sec:level1} introduction}

To bind Fenrir, the fearsome wolf in Norse mythology, the gods forged the strongest chain by combining the softest and most elusive elements of the universe---the sound of a cat's footsteps, the breath of a fish, and the spittle of a bird. Unexpected properties can indeed emerge, almost mysteriously, from systems made up of discrete elements with vastly different characteristics. Examples span a wide range of scales: the anomalous physical properties in different phases of water originate from the orientation-dependent hydrogen-bonding interactions between its molecules \cite{Pettersson2015}, pacemaker cells synchronize to produce a single cardiac impulse \cite{Peskin1975}, and swarm of bacteria or ants perform complex functions as a unit \cite{Kronauer2022, Peng2021_collective}. To formalize, predict, and ultimately control the emergent behaviors of multi-element systems, we need not only to understand the interactions between the constituents but also to establish the inherent link between these microscopic interactions and macroscopic material properties \cite{Mataric1993}. Composites and metamaterials offer a viable system to accomplish such a challenging task, allowing us to engineer new materials with desired properties by tailoring the structures and interactions of their basic building blocks \cite{Milton2019}. 

\begin{figure*}[t!]
\centering
\includegraphics[width=0.8\textwidth]{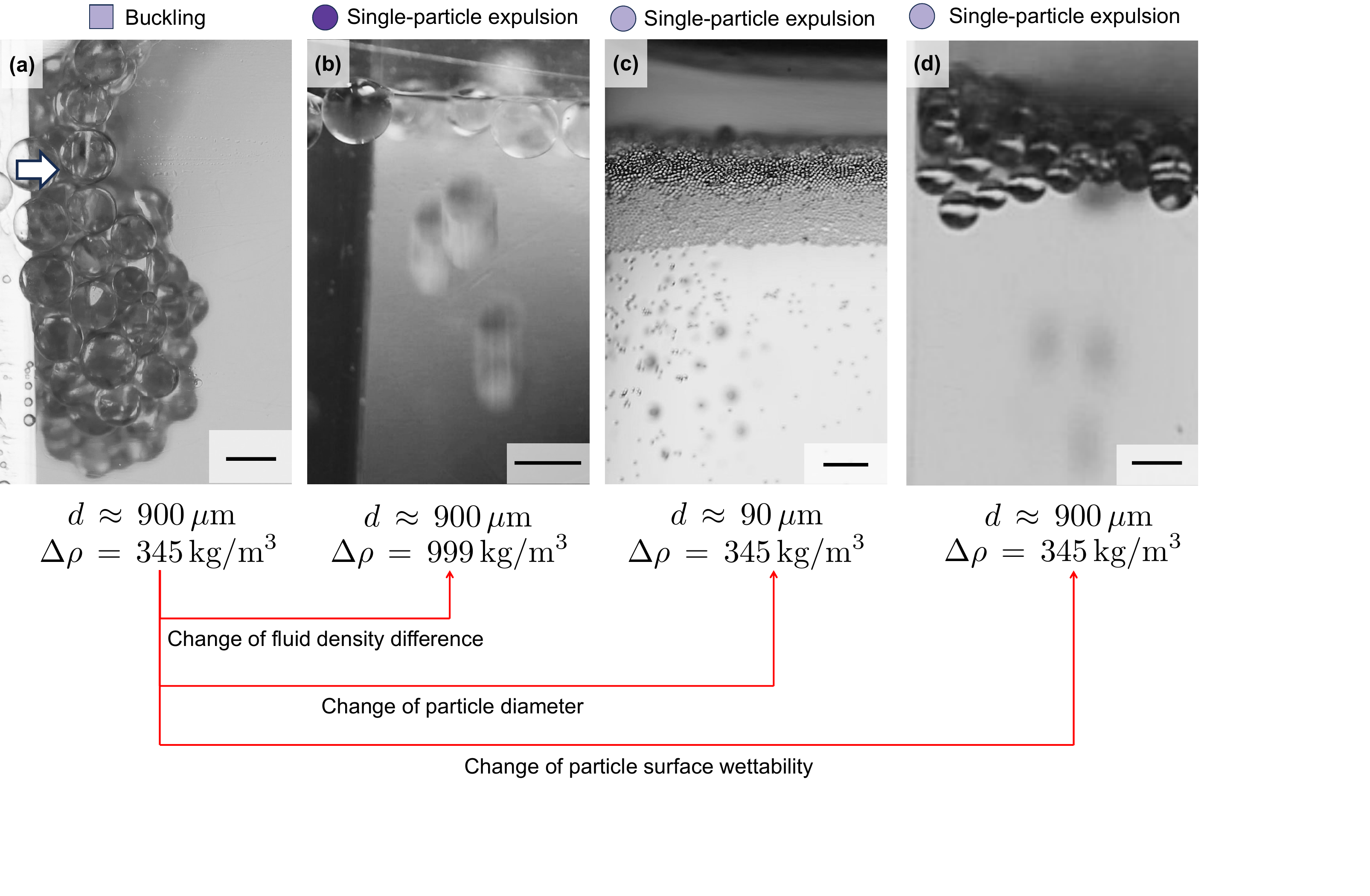}
\caption{Failure of particle rafts under compression. (a) Side-view of a granular raft that fails through buckling (denoted by a square at the top) at a hexane-water interface ($\Delta\rho\,=\,345\,\text{kg/m$^3$}$) during compression in a trough. The fold location, in most cases, is near the moving wall (Appendix A, section 2). The white arrow shows the compression direction for all the images. (b) A raft made of similarly sized particles fails via single-particle expulsion (denoted by a circle) at an air-water interface ($\Delta\rho\,=\,999\,\text{kg/m$^3$}$). While keeping the interfacial fluid combination the same as the hexane-water interface, buckling to particle-expulsion transition can also be obtained (c) by reducing the particle diameter from $d\,\approx\,900\,\mu\text{m}$ to $d\,\approx\,90\,\mu\text{m}$ or (d) by making the particle surface more hydrophilic. All the scale bars are 1 mm.}
\label{fig:schema}
\end{figure*}

As perhaps the simplest composite material, particle rafts composed of otherwise cohesionless particles behave collectively as an elastic membrane once aggregated at a fluid-fluid interface \cite{Vella2005}. These remarkable membrane-like properties of particle rafts are on display in Fig.\,\ref{fig:schema}(a), where rafts form out-of-plane wrinkles and fold under compression \cite{Vella2004a, Monteux2007, Leahy2010a, Protiere2017, Druecke2023} (Movies S1 and S3 \cite{supp}), similar to the behavior of a compressed elastic sheet \cite{Huang2007, Pocivavsek2008a}. In contrast, the granular nature of rafts is highlighted in Fig.\,\ref{fig:schema}(b), where, instead of a raft-scale buckling, compression results in individual particles being expelled from the interface \cite{Razavi2015, Druecke2023} (Movie S1 \cite{supp}). The dual nature of particle rafts exhibiting either membrane-like \cite{JambonPuillet2017} or granular characteristics \cite{Cicuta2009} underpin their numerous technological applications ranging from the stabilization of fluid-fluid interfaces \cite{Binks2005, Abkarian2013, Pickering1907, Reiss2003} and formation of bijels \cite{Stratford2005_bijels, Cates2008_bijels, Bai2015_bijels} to the controlled release of particles in biomass conversion and gas storage \cite{Resasco2010, Garbin2015}. A deeper understanding of the transition between these mutually exclusive behaviors of particle rafts could allow for more precise design and control of their dynamics tailored to specific applications.

Mapping the particle-scale features and interactions to the emergent raft dynamics is far from trivial. Even though changes in particle and fluid properties demonstrably affect the behavior of the particle-laden interfaces \cite{Vella2004a, Bordacs2006, Planchette2012, Garbin2015, Razavi2015, Druecke2023}, a complete understanding of the physical origin of different raft behaviors is still missing \cite{Vella2004a, Protiere2017, Gu2018, Druecke2023}. Existing continuum models rely on empirical bending moduli to explain the elastic nature of rafts \cite{Vella2004a, Protiere2017} and fail to capture their granular characteristics. Here, in support of our systematic experimental observations, we present a first-principle model that incorporates the microscopic contact line dynamics of particles governed by a set of physicochemical parameters of both the particles and the fluids. The dual nature of particle rafts in our model arises from the competition between the energy required to detach a single particle and the energy needed to buckle the entire interface. By minimizing the system's energy, the model naturally produces a term analogous to the ad-hoc bending modulus proposed in previous studies, thus revealing the microscopic basis of the rafts' bending stiffness. Finally, guided by the model prediction on the microscale-macroscale relationship, we demonstrate the capability of engineering raft behaviors by altering the contact line dynamics of individual particles at the fluid interface.

\linespread{1}

\section{\label{sec:level2} Understanding the failure mode of particle rafts}
\subsection{\label{sec:level2a}Three features determining how a particle raft fails under compression}

We investigate the dynamics of rafts in bi-axial and linear compression experiments using a funnel and a trough, respectively [Appendix A, Fig.\,\ref{fig:compression}, Movies S1 and S2 \cite{supp}]. Results from these methods are used interchangeably as they are quantitatively similar. Our experimental observations establish the essential role of fluid densities ($\Delta\rho\,=\,\rho_1-\rho_2$), particle diameter ($d$), and particle surface wettability in dictating the raft failure modes during compression (Movie S2 \cite{supp}). Rafts predominantly fail via particle expulsion for higher $\Delta\rho$, as shown by the transition from buckling at a hexane-water interface [Fig.\,\ref{fig:schema}(a)] to particle expulsion at an air-water interface in Fig.\,\ref{fig:schema}(b). Moreover, under the same fluid combination, when the particle diameter is reduced [Fig.\,\ref{fig:schema}(c)] or the particle surface is made more hydrophilic [Fig.\,\ref{fig:schema}(d)], a similar transition from buckling to particle expulsion can be observed. Although the effects of fluid densities and particle diameter on raft properties and behaviors have been reported previously \cite{Vella2004a, Planchette2012, Protiere2017, Druecke2023}, the exact role of the surface wettability on raft dynamics remains an open question \cite{Planchette2012, JambonPuillet2017}, which leads to an incomplete understanding of particle raft behaviors.  

\begin{figure*}[t!]
\centering
\includegraphics[width=\textwidth]{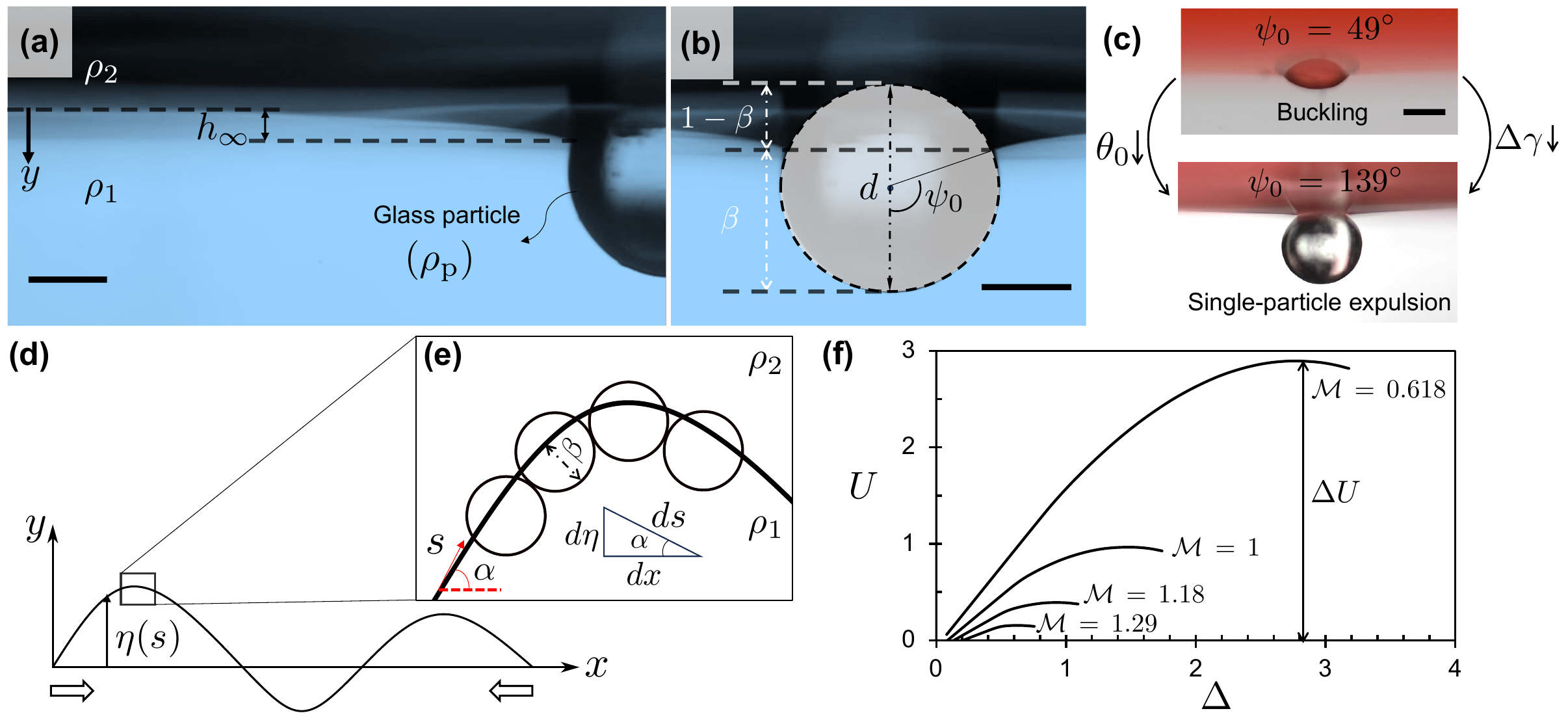}
\caption{A model for the particle raft failure. (a) Particle with density $\rho_\text{p}$ floating at a fluid-fluid interface with densities $\rho_1$ and $\rho_2$. All the distances are measured from the undisturbed horizontal interface far from the particle. The solid-fluid contact line is $h_\infty$ distance away. (b) Particle equilibrium position at an interface is given by $\psi_0$, the angle between the center line of the particle and the contact line. $\beta$ is the fraction of the particle below the contact line within the lower fluid. (c) When the particle position at a hexane-water interface changes from $\psi_0\,=\,49^{\circ}$ to $\psi_0\,=\,139^{\circ}$, corresponding raft failure mode changes from buckling to single-particle expulsion. (d,e) The continuum model of the raft is a one-dimensional array of spheres constrained in the ($x, y$) plane between two fluids with densities $\rho_1$ and $\rho_2$. Raft deformation is denoted by the vertical distance, $\eta(s)$, where ($s,\alpha$) are the intrinsic coordinates used for the energy calculations and equilibrium equation. (f) Variation of the dimensionless total raft energy $U$ with raft compression $\Delta$, for given $\mathcal{M}$. $\Delta U$ denotes the energy barrier for buckling the particle raft. All the scale bars are $500\,\rm{\mu m}$.}
\label{fig:static}
\end{figure*}

To understand the relationship between particle surface wettability and raft failure modes, we visualize the individual particles and extract their static equilibrium position on a given fluid surface as shown in Fig.\,\ref{fig:static}(a). We define the static equilibrium position by the angle $\psi_0$ between the particle centerline and the three-phase contact line [Fig.\,\ref{fig:static}(b)]. The fraction of the particle below the three-phase contact line is given by $\beta\equiv (1/2)(1-\cos{\psi_0})$, which can vary from $\beta=1$ (full submersion into fluid 1) to $\beta=0$ (zero submersion). The location of the contact line can also be denoted by the length $h_\infty$ measured from the undisturbed fluid-fluid interface. As expected, the surface wettability of the particles plays an important role in determining their static equilibrium positions at a fluid-fluid interface. When a cleaned glass bead of diameter $\approx\,900\,\rm{\mu m}$ is placed on a hexane-water interface (Appendix A, section 2), the majority of the particle stays within the hexane phase with $\psi_0\,=\,49^{\circ}$ [Fig.\,\ref{fig:static}(c) top panel]. In contrast, when uncleaned particles of the same size are placed at the hexane-water interface, $\psi_0$ is increased to $139^{\circ}$ [Fig.\,\ref{fig:static}(c) bottom panel]. The hydrophilicity of the uncleaned particles is probably a result of soda-lime glass surfaces acquiring a micrometric thin layer of water film in a humid environment \cite{Caurant2020}. We observe that rafts made of clean $d\,\approx\,900\,\rm{\mu m}$ particles fail via buckling, while rafts made of hydrophilic particles of the same size fail through single-particle expulsion. To further test if the static equilibrium position of the particles at the interface effectively controls the raft dynamics, we also increase $\psi_0$ of a clean particle by adding surfactant in the water phase. As surface tension $\gamma$ is lowered from $43\,\rm{mN/m}$ to $14\,\rm{mN/m}$, $\psi_0$ increases to $137^{\circ}$. Again, the raft made at this modified interface fails by particle expulsion, demonstrating the important role of the equilibrium position of the particle in determining the raft dynamics under compressive loading.

\linespread{1}

\subsection{\label{sec:level2b} Battle of two energies}

Our experimental observations require a model that can simultaneously capture the granular and elastic nature of the raft and explain the effects of three key physical parameters---fluid densities, particle size, and equilibrium particle position---on the failure of the raft. Rather than performing grain-scale simulations that are computationally expensive, we develop two distinct simplified models: one that treats the particle raft as an elastic sheet and another that isolates a single particle on the interface. Using the two models, we compare the energies required for each failure mode---system-wide buckling and particle expulsion, as a function of the key physical parameters. We hypothesize that the competition between the two energies effectively determines the raft failure.

First, we model the raft as a 1D continuous sheet by considering an array of spherical particles in the intrinsic coordinates $(s,\alpha)$, where $s$ is the arclength and $\alpha$ is the angle between the line tangent to the array and the horizontal axis [Fig.\,\ref{fig:static}(d) and Fig.\,\ref{fig:static}(e)]. Previous studies have successfully used the beam equation to model the raft with an effective bending modulus $B$ that is determined empirically \cite{Monteux2007, Varshney2012, Planchette2012, Lagubeau2014}. Instead of relying on an ad-hoc bending modulus, we derive the equation for the shape of the compressed particle raft via energy minimization at a given compression level $\Delta$, an approach similar to that used to describe a floating elastic sheet \cite{JambonPuillet2016a} (Appendix B).

The total energy $U$ of the raft includes the potential energies of the particles $U_p$ and of the fluids $U_f$ and the interfacial energies of the system $U_i$ (Appendix B, Eq.\,\ref{eq:raft}), which we non-dimensionalize with $Kd/\ell_e$. Here, $K\equiv (3\pi/8)\gamma d^3 \phi_s\sin^4{\theta_\text{eq}}$, where $\phi_s=1/d$ is the number of particles per unit length of a jammed raft, while $\ell_e\equiv(K/\Delta\rho g)^{1/4}$ is the characteristic length scale. We set the value of the equilibrium contact angle, $\theta_\text{eq}$, based on the independent wettability measurements with a glass slide (Appendix A, Fig.\,\ref{fig:angles}). Notably, $U_i$ represents the energetic cost associated with the curvature of the raft, which is equivalent to bending. By differentiating the interfacial energy with the curvature, we obtain a bending modulus $B\equiv (3\pi/8)\gamma d^3 \phi_s\sin^4{\theta_\text{eq}}$, which has the same expression as $K$. This bending modulus, obtained from the raft energy minimization, has the same dependence on particle diameter, surface wettability, and interfacial tension as that adopted phenomenologically in other continuum models \cite{Vella2004a, Planchette2012}. The length scale $\ell_e$ is equivalent to $(B/\Delta\rho g)^{1/4}$, the universal expression of wrinkle wavelength for compressing a thin elastic sheet on an elastic foundation \cite{Pocivavsek2008a, Cerda2003}. 

Once the shape of the raft $\alpha(s)$ is computed numerically (Appendix B, Eq.\,\ref{eq:shape_alpha}), we calculate the total dimensionless energy $U^*=U/(Kd/\ell_e)$ of the compressed raft as 
a function of the degree of compression $\Delta^*=\Delta/\ell_e$, which is controlled by a dimensionless parameter:
\begin{equation}
    \mathcal{M}\equiv\frac{\pi d^2\phi_s}{6 \ell_e}\left[\frac{\rho_{\rm p}-\rho_2}{\rho_1-\rho_2}-\beta\right].
    \label{def:M}
\end{equation}
$\mathcal{M}$ denotes the effective dimensionless weight of the particle layer \cite{JambonPuillet2016a}. However, in contrast to the existing continuum models of rafts \cite{Protiere2017}, $\mathcal{M}$ explicitly depends on the equilibrium position of particles at the fluid interface via $\beta$. Importantly, for given $\mathcal{M}$, the raft energy $U$ varies non-monotonically with $\Delta$ [Fig.~\ref{fig:static}(f)], which provides an energy barrier preventing the buckling of the interface. Thus, the energy required to buckle an interface is determined by the difference between the maximum $U$ and $U(\Delta\approx 0)$, denoted as $\Delta U$. Note that we have dropped the asterisk used to denote the dimensionless variables for brevity. 

Second, we compute the total work required to expel a particle from its initial equilibrium position at the interface. For simplicity, we neglect the effects of the neighboring particles and consider a spherical particle on a bare interface that is pushed quasi-statically into the lower fluid. The net downward force to detach the particle can be computed from the force balance, where the downward force $F$ and the weight of the particle $F_g$ must be balanced by the surface tension force $F_\gamma$ acting at the three-phase contact line and the buoyancy force $F_b$ \cite{Vella2006}. Computing $F_\gamma$ and $F_b$ requires simultaneously solving for the shape of the fluid-fluid interface $h$ via the Young-Laplace equation, the details of which are included in Appendix C.

\linespread{1}
\begin{figure*}[t!]
\centering
\includegraphics[width=\textwidth]{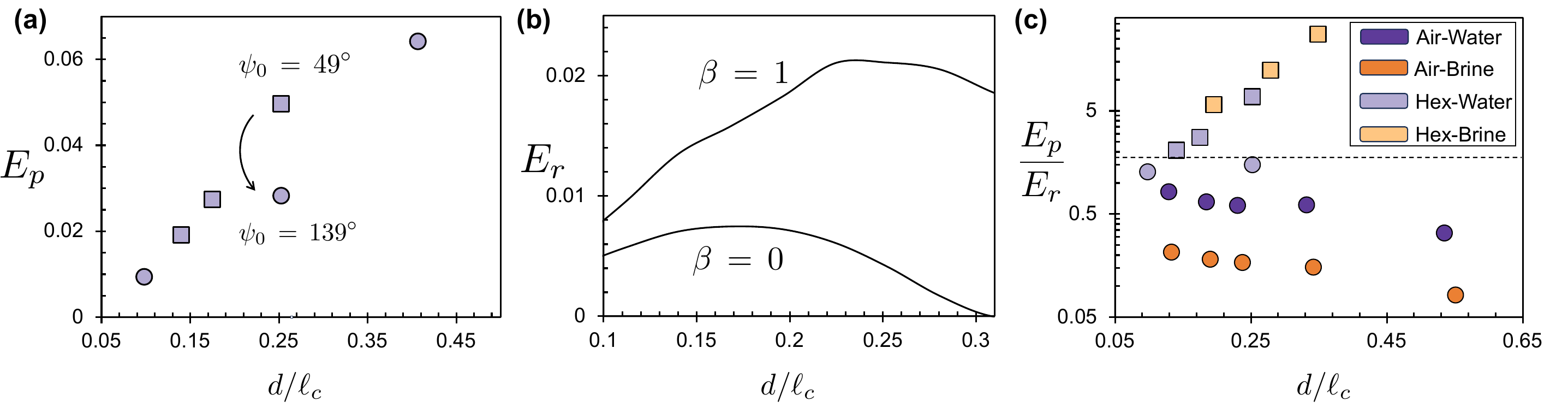}
\caption{Comparing the energies to fail a raft. (a) Non-dimensional energy $E_p$ to detach a particle from the hexane-water interface. $E_p$ monotonically increases with non-dimensional particle size $d/\ell_c$. The change in equilibrium position for cleaned ($\psi_0\,=\,49^{\circ}$) and unclean ($\psi_0\,=\,139^{\circ}$) particles reflects in the decrease in $E_p$. Symbols for the energy values signify the corresponding raft failure mode. (b) Non-dimensional energy to buckle a particle-laden hexane-water interface, $E_r$ changes non-monotonically with particle size $d/\ell_c$. For any particle position $0\,<\beta\,<\,1$, the energy to buckle a particle-laden hexane-water interface would fall between the two lines in the plot. (c) Phase map of different failure modes based on the energy ratio $E_p/E_r$ and scaled particle size $d/\ell_c$. For $E_p/E_r\,>\,1$, the energy to expel a particle from the interface is more than the energy to buckle the interface. The experimental observations of buckling are all placed above the dashed horizontal line at $E_p/E_r\,=\,1.8$.}
\label{fig:model}
\end{figure*}

Once the downward force $F$ and the instantaneous interfacial shape $h$ are known as a function of the vertical position $y$ [Fig.\,\ref{fig:static}(a)], we estimate the work required for particle expulsion $E_p$ by integrating $F$ over $y$. We non-dimensionalize the lengths with the capillary length, $\ell_c\equiv\sqrt{\gamma/\Delta\rho g}$, and the force by $\Delta\rho g\ell_c^3$, so that $E_p^*\,=\,E_p/(\Delta\rho g\ell_c^4)$. As before, we drop the asterisk corresponding to the dimensionless variables for brevity. We separate the particle expulsion into two stages to account for the contact angle hysteresis. Stage (\textit{i}): The contact angle increases from the initial value $\theta_0$ to the advancing contact angle $\theta_a$, while remaining pinned at $\psi_0$. Stage (\textit{ii}): Once the advancing angle is reached, the contact line moves with constant $\theta_a$. Hence,
\begin{equation}
E_p=\int_{y_c(\theta_0)}^{y_c(\theta_a)} \,F \,{\rm d} y \bigg|_{\psi=\psi_0}+\int_{y_c(\psi_{a})}^{y_c(\psi_d)} F \,{\rm d} y \bigg|_{\theta=\theta_a},
\label{eq:particle_energy_main}
\end{equation}
where $y_c(\theta,\psi)=h_\infty(\theta,\psi)-(d/\ell_c)\cos{\psi}/2$ is the instantaneous position of the particle center, and $\psi_a$ and $\psi_d$ denote the contact line positions at depinning (i.e., when $\theta\,=\,\theta_a$) and upon detachment, respectively. The detachment of the particle from the interface is set when $F$ reaches a maximum value (Appendix C, Fig.\,\ref{fig:Ftoyc}).

Fig.\,\ref{fig:model}(a) is a representative plot at the hexane-water interface that shows how the particle expulsion energy, $E_p$ varies with particle diameter $d$ rescaled by the hexane-water capillary length $\ell_c\,=\,3.56\,\text{mm}$. To calculate $E_p$ from Eq.\,(\ref{eq:particle_energy_main}), the value of $\psi_0$ is obtained from the static equilibrium measurements, and $\theta_0$ is found by solving the Young-Laplace equation with the initial static position of the particle at the interface ($y_c$, $\psi_0$) as the input. In addition, we use $\theta_a$ from the dynamic particle removal experiments where we attach the particle at the end of a needle and quasi-statically push it into the lower fluid (Appendix A, section 2, Fig.\,\ref{fig:dynamic}). The results show that the particle expulsion energy increases monotonically with the particle size. At $d/\ell_c\,=0.25$, making particle surface more hydrophilic lowers its equilibrium position ($\psi_0\,=\,49^{\circ}$ to $\psi_0\,=\,139^{\circ}$) and the energy required to remove the particle from the interface also decreases as expected. 

Finally, to compare the required buckling energy for the particle raft with the expelling energy for a single particle, we rescale $\Delta U$ from the continuum model to be consistent with the non-dimensionalization of $E_p$ in the particle model and label it as $E_r$. We again plot $E_r$ as a function of $d/\ell_c$ for the representative case of a hexane-water interface in Fig.\,\ref{fig:model}(b). We observe that the energy to buckle an interface is higher when the particles are completely within hexane ($\beta\,=\,0$) compared to when they are fully submerged in water ($\beta\,=\,1$). At these extrema, the variation of $E_r$ with particle size is non-monotonic, i.e., it increases till a certain size but then drops to zero for particles more than a millimeter in diameter. As the particle positions will be in the range $0\,<\,\beta\,<\,1$, the energy to buckle a particle-laden hexane-water interface would fall between the extrema. 

\linespread{1}
\begin{figure*}[t!]
\centering
\includegraphics[width=\textwidth]{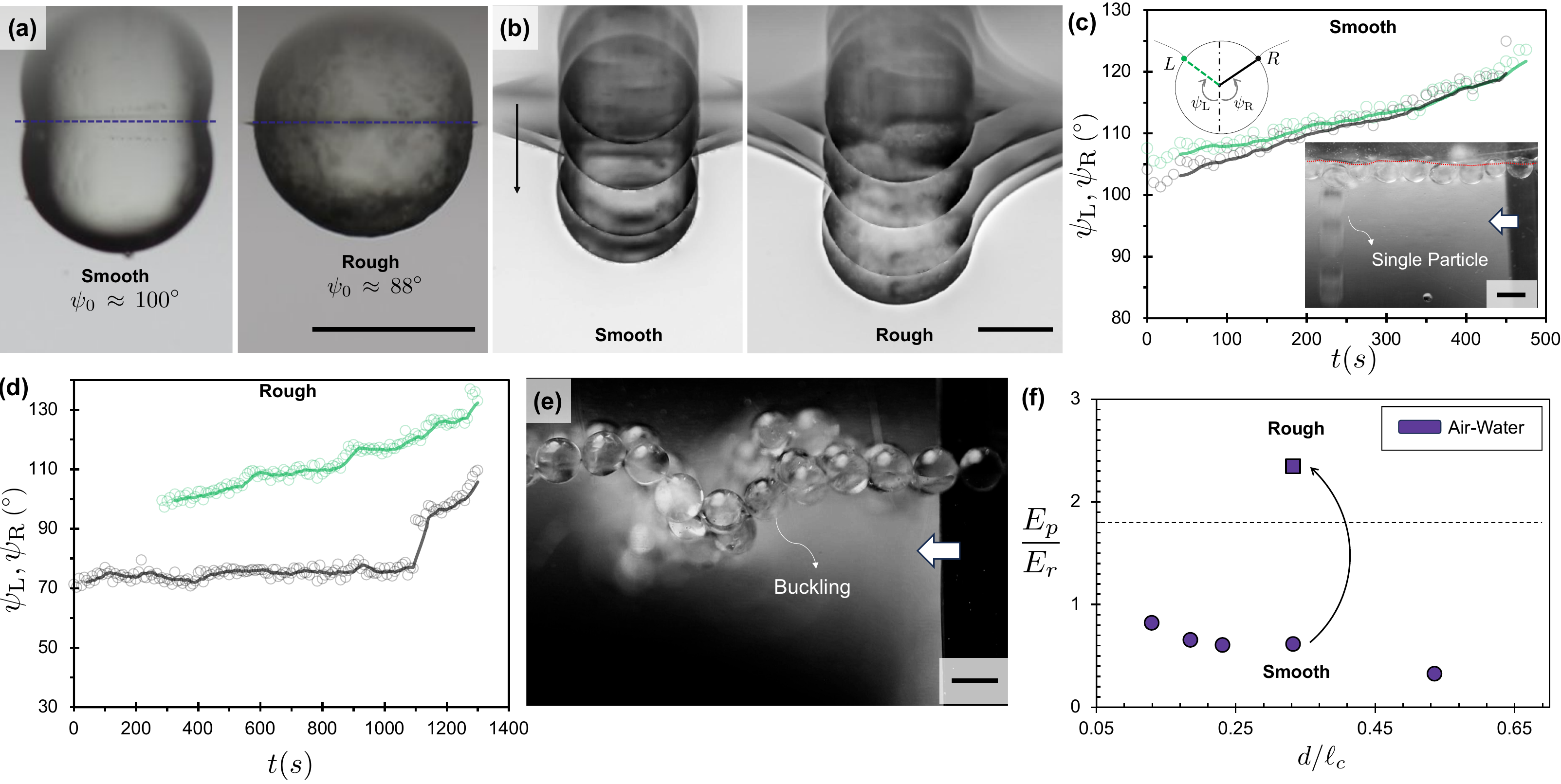}
\caption{Controlling raft dynamics. (a) Static equilibrium positions of a similarly-sized smooth and rough glass bead at the air-water interface are comparable. (b) Superimposing four frames at a specific time interval from the dynamic particle expulsion experiments at the air-water interface shows the difference between a smooth and rough glass bead. The contact line slides along the particle surface, maintaining a near-constant contact angle as the smooth particle moves downward (shown by the black arrow). In contrast, on the rough particle, the pinning of the contact line is evident from the significant deformation of the meniscus and the resulting change in the contact angle. (c) (Upper inset) The contact line is tracked with the right (`R', black) or left (`L', green) points on a particle as it pierces the interface. The location of the contact points is measured by the angle $\psi$ on either side. The time series plot of contact points on a smooth particle shows a continuous motion of the contact line. (Lower inset) A raft made of smooth particles fails via single-particle expulsion. The dotted red line shows the interface. (d) ``Stick-slip'' motion of the contact line is evident when a similar-sized rough particle is pushed through the air-water interface. (e) The raft made of rough particles fails via buckling at the air-water interface. (f) The model prediction shows that geometric heterogeneities on a particle surface increase the energy ratio $E_p/E_r$ confirming the experimental observation. All the scale bars are $1\,\rm{mm}$.}
\label{fig:roughness}
\end{figure*}
 
We hypothesize that the raft dynamics are effectively determined by a competition between the energy terms $E_p$ and  $E_r$ and compute their ratio $E_p/E_r$ for all our experiments. Our experimental results for four different fluid-fluid interfaces, particle sizes, and surface wettabilities can then be compiled into a phase map of raft failure modes spanned by $E_p/E_r$ and non-dimensional particle size, $d/\ell_c$ [Fig.~\ref{fig:model}(c)]. The dashed horizontal line at $E_p/E_r\,>\,1$ demarcates the boundary between the particle expulsion and buckling failure as observed in the experiments. The consistently low values of the energy ratio for both air-water and air-brine interfaces, irrespective of $d$, agree well with our experiments, where we have always seen single-particle expulsion at liquid-air interfaces. This is in contrast to a previous theoretical model \cite{Druecke2023}, which predicted that increasing the particle size in a raft at the air-water interface would result in a transition from particle expulsion to buckling. The map also incorporates the effect of particle surface wettability in the raft dynamics, an important factor either ignored or rendered insignificant in existing continuum models of particle rafts \cite{Vella2004a, Planchette2012, Druecke2023}. Using a more wettable particle surface would result in a sharp drop in the value of the energy ratio, which explains the observation of single-particle expulsion at the hexane-water interface with $d\,\approx\,900\,\mu\text{m}$ particles. Although not included in our phase map of Fig.~\ref{fig:model}(c), our model predicts that for mineral oil-water interfaces, the energy to buckle the interface is negligible, rendering a very high value for the energy ratio $E_p/E_r$. This is perfectly in tune with the experiments where irrespective of particle size ($d\,\approx\,90\,\mu$m to $900\,\mu$m), buckling of the interface is the general mode of raft failure. For similar reasons, experimental results at the hexane-brine interface with $d\,\approx\,900\,\mu$m and $d\,\approx\,1450\,\mu$m are also not included in the phase map, despite the model results agreeing with the experimental observations (Fig.\,\ref{fig:phase-map}).

A few points warrant further discussion. First, our single-particle calculation provides a lower bound on $E_p$. Treating particle motion quasi-statically, the model ignores the effect of viscous drag and the energy associated with the slipping of the contact line when the particle moves through the subphase fluid. These idealizations are justified in our study because, for the fluid interfaces used in the phase diagram of Fig.\,\ref{fig:model}(c), the capillary number associated with the particle expulsion experiments is $\sim\,10^{-5}$ signifying negligible viscous effects. Second, Fig.\,\ref{fig:static}(f) shows that the energy barrier to buckling decreases with increasing $\mathcal{M}$ and eventually becomes zero. Physically, as $\mathcal{M}$ measures the effective weight of the particle raft relative to the effective restoring force (due to surface tension and buoyancy), with increased raft weight or decreased restoring force, the external energy required to buckle a raft would decrease. Thus, our model predicts that the raft will collapse due to its own weight beyond some critical $\mathcal{M}$, whose value could be better quantified via a full three-dimensional raft model, outside the scope of this work. Finally, our model predicts non-zero energy input for expelling individual particles even when they float at the interface in a metastable state \cite{vella2006load}. Future modifications of the model should also consider the effects of metastable floating conditions. Despite the approximations ignoring these subtle effects, our model accurately describes the macroscopic raft dynamics in experiments with a wide range of particle and fluid parameters, as evident from the phase map in Fig.\,\ref{fig:model}(c) and Fig.\,\ref{fig:phase-map}. 

\section{\label{sec:level3} Engineering the failure mode of particle rafts}

In addition to capturing the role of contact angle or static equilibrium position of particles, our model incorporates the effects of contact angle hysteresis. As per Eq.\,(\ref{eq:particle_energy_main}), the energy to quasi-statically remove a particle from a fluid interface can be modified through the ``stick-slip'' motion of the contact line where the contact line initially remains stuck and suddenly depins to move between a series of pinning sites \cite{Quere2008}. We hypothesize that if such a change is achieved without any appreciable change to the raft buckling energy $E_r$, the corresponding raft dynamics in compression will also change. Establishing this link between the three-phase contact line motion and the overall raft dynamics provides new insight, enabling us to modify the raft behavior through particle-scale manipulations.

As surface roughness can act as pinning sites \cite{Snoeijer2013, Robbins1990, Weinan2011}, we roughen glass beads with $d\,\approx\,900\,\mu\text{m}$ by rolling them between two pieces of sandpaper. Although the equilibrium position of an unaltered `smooth' glass bead is similar to that of a `roughened' glass bead at an air-water interface [Fig.\,\ref{fig:roughness}(a)], the quasi-static particle expulsion experiments highlight their differences. Fig.\,\ref{fig:roughness}(b) compares the contact line motion on a smooth and a rough particle as it is expelled from an air-water interface  (Movie S4 \cite{supp}). The images are created by superimposing four frames from respective particle expulsion experiments at specific intervals. For the smooth particle, the contact line moves along the particle surface, maintaining a constant contact angle. By contrast, contact line pinning is evident on the right side of the rough particle. We quantitatively compare the contact line motion between the two particles by tracking the points `L' and `R' on the particle surface, shown by the green and black dots in the inset of Fig.\,\ref{fig:roughness}(c). As evident from the superimposed contact line motion for the rough particle, tracking one point on either side of the particle ensures the spatial variability of the pinning locations is captured \cite{Jung2017}. The position of the contact line is denoted by the angles $\psi_\text{L}$ and $\psi_\text{R}$ corresponding to points `L' and `R', respectively. The result of tracking a smooth particle is plotted in Fig.\,\ref{fig:roughness}(c). The green and black lines indicate the time-averaged movement of the contact line. Their almost constant slope signifies an uninterrupted contact line motion as the particle pierces the air-water interface. Rafts made of these particles at the air-water interface always show single-particle expulsion, as depicted in the chronophotograph from the experiment at the inset of Fig.\,\ref{fig:roughness}(c). A stark difference is observed in the dynamic behavior of rough particles in Fig.\,\ref{fig:roughness}(d). The variation of $\psi_\text{R}$ signifies that the contact line remains stuck for a long period and suddenly jumps to a higher advancing angle. The classic ``stick-slip'' motion on a roughened surface is evident in the variation of $\psi_\text{L}$. When these rough particles are used in a raft at an air-water interface, the raft fails via buckling instead of single-particle expulsion [Fig.\,\ref{fig:roughness}(e)]. Thus, our experiments demonstrate how to engineer macroscopic raft behaviors by changing microscopic particle-scale features (Movie S5 \cite{supp}).

To further explain our observation, we assume a model rough sphere where the local advancing contact angle is a sinusoidal function of the contact line position $\psi$ given by $\theta=\theta_a+\theta_b\sin(\pi(x-(\psi-1)/2))\left(H(\psi-1)-H(\psi+1)\right)$. Here, $\theta_b$ is the change of advancing contact angle due to a heterogeneous surface, and $H$ is the Heaviside function, which ensures that the change of contact angle only occurs within $\pm 1^\circ$ of $\psi$. We set $\theta_b=10^\circ$ and select 5 random $\psi$ values as the pinning locations along the particle surface. The process of pinning and depinning of the contact line inevitably increases the particle expulsion energy, $E_p$ (Eq.\,\ref{eq:particle_energy_main}). As the interfacial properties ($\gamma$, $\Delta\rho$) along with the particle equilibrium position ($\beta$) and diameter ($d$) remain unchanged, the raft buckling energy $E_r$ for the smooth and rough particle is the same in our model. This pushes the energy ratio $E_p/E_r$ for rough particles above the critical transition, indicated by the arrow in Fig.\,\ref{fig:roughness}(f). 

\section{\label{sec:level4} Conclusions}

Similar to micromechanical analysis for predicting the anisotropic response of composites \cite{Buryachenko2007}, our energy model predicts macroscopic raft dynamics in compression by analyzing static equilibrium position and quasi-static motion of individual particles at the fluid interface. We have identified a critical criterion that determines the raft failure mode during compression: if the energy required to detach a particle from the interface exceeds the energy to buckle a particle-laden interface, the raft would preferably fail by buckling. Informed by our model prediction, we show that the energies and, consequently, the failure mode of rafts can be controlled not only by particle size, surface wettability, or physicochemical modifications of the interface but also through surface textures. 

Although our study focuses only on rafts with larger particles, the theory presented here should be generally applicable to rafts of a wide range of particle sizes. The dynamics of colloidal particle-laden interfaces, both in the planar \cite{Bordacs2006, Razavi2015a} and spherical geometry \cite{Abkarian2007, Pitois2015}, are often analyzed based on the area- or volume-dependent surface pressure variations, rather than the minimization of system energy. However, the surface pressure is extremely sensitive to the measurement methods \cite{Kumaki1988}, resulting in disparate conclusions about the general mechanism of colloidal raft destabilization \cite{Aveyard2000a, Tolnai2001}. Our micro-to-macro energy model overcomes this difficulty and provides a unifying theory for predicting and controlling the macroscopic behavior of generic particle rafts. From the engineering point of view, the phase map presented here provides a useful framework to control the stability of particle-laden interfaces, which would be relevant in applications requiring targeted fluid or particle delivery \cite{Garbin2015} and arrest or trigger ripening in foam and emulsion systems \cite{Vermant2017}. 

\begin{acknowledgments}
This work was supported primarily by the National Science Foundation through the University of Minnesota MRSEC under Award No. DMR-2011401, and through CBET-2032354 and CMMI-2042740. We are particularly grateful to Etienne-Jambon Puillet for providing insightful comments and generously sharing his methodologies and codes. R.M. also wants to thank Zak Kujala for suggestions, help in the experimental setup, and stimulating discussions.
\end{acknowledgments}

\appendix
\section{Materials \& Methods}

\subsection{Particles and Fluids}

The particles used in this study are made of soda-lime glass (Ceroglass, $\rho_\text{p}\approx\!2500$\,kg/m$^3$). They come in different ranges of diameters: the minimum diameter ($d$) range is $80\,\mu\text{m}-100\,\mu\text{m}$ and the maximum is $d\,=\,1250\,\mu\text{m}-1650\,\mu\text{m}$. Actual particle diameters are measured for the static equilibrium and dynamic experiments with single particles. For the plots in Fig.\,1 and Fig.\,3, we have used the mean value of the diameter range with error bars as mentioned in their respective packages. Thus, we have 8 different mean particle diameters for rafts: $90\pm20\,\mu\text{m}$, $150\pm50\,\mu\text{m}$, $250\pm50\,\mu\text{m}$, $350\pm50\,\mu\text{m}$, $500\pm100\,\mu\text{m}$, $625\pm125\,\mu\text{m}$, and $900\pm100\,\mu\text{m}$. Apart from these soda-lime glass beads, we have also used borosilicate glass beads (Mo-Sci Corp, $\rho_\text{p}\approx\!2200$\,kg/m$^3$, $d\,=\,100\pm20\,\mu\text{m}$) for a few experiments. 


Particles are prepared in three different ways to introduce the variability in the surface wettability of the particles: 1) using the particles as is, 2) rinsed, and 3) particles passed through a Piranha solution and sonicated for 10 minutes. The non-rinsed particles are usually very hydrophilic, as evident in their lower equilibrium position at an air-water interface. When these particles are first rinsed successively in a soap solution, regular tap water, and then distilled water followed by drying in an oven for 4 hrs at 200$^\circ$C, their contact angle in water increases. The Piranha-treated particles are extremely hydrophilic as we couldn't form a raft with them.

Three different liquids are used as the heavier liquid phase in the experiments. These are distilled water ($\rho\,\approx\,1000$\,kg/m$^3$, $\gamma\,\approx\,72$\,mN/m), aqueous glycerol solution ($\rho\,\approx\,1210$\,kg/m$^3$, $\gamma\,\approx\,58$\,mN/m), and brine ($\rho\,\approx\,1195$\,kg/m$^3$, $\gamma\,\approx\,81$\,mN/m). The upper lighter fluids forming the fluid-fluid interface are air or hexane ($\rho \approx\,655$\,kg/m$^3$, $\gamma\!\approx\,43$\,mN/m). The values of density and surface tension in air for all these liquids are given within the parenthesis. All the liquids except distilled water are from Sigma-Aldrich with 95-99\,\% purity. In the experiments where the surface tension of the water phase is lowered, we use 1.33 mM and 2.67 mM aqueous concentrations of sodium dodecyl sulfate (SDS, Sigma Aldrich).

\subsection{Experimental design and protocol}

\begin{figure*}[!ht]
\centering
\includegraphics[width=0.7\textwidth]{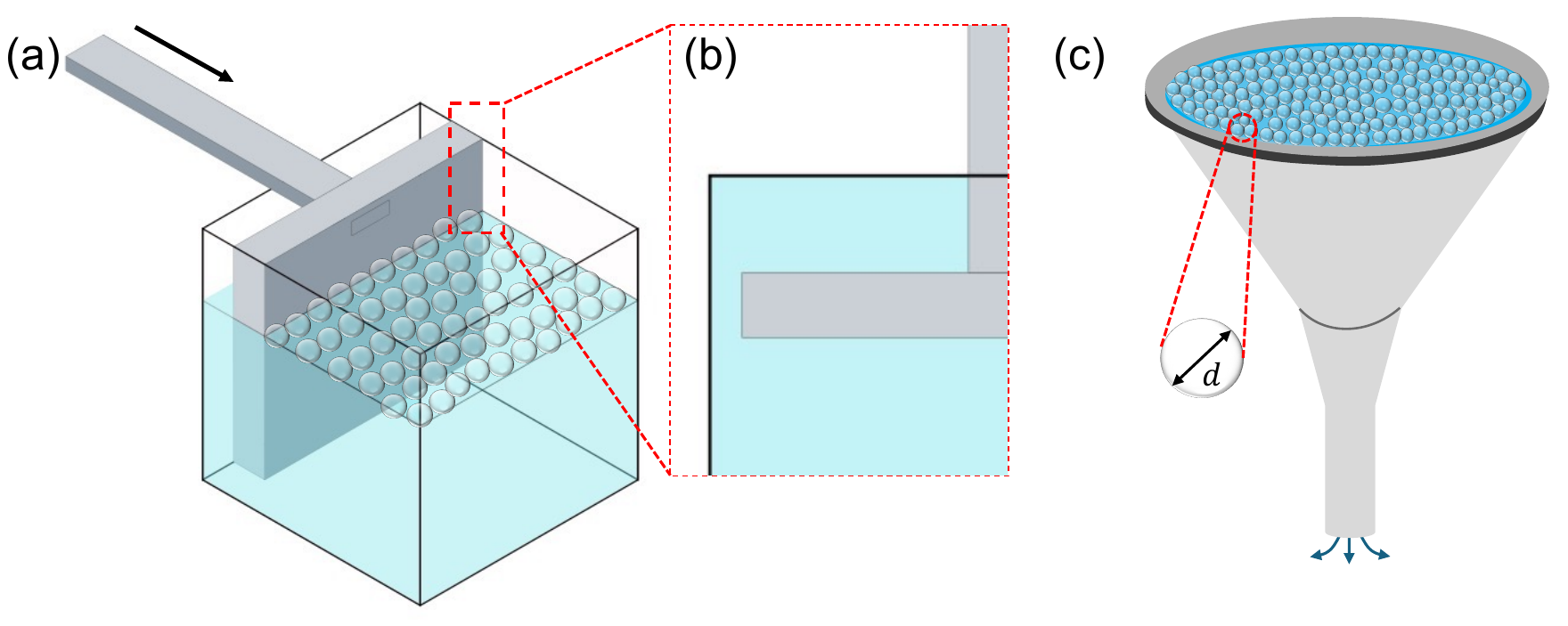}
\caption{Compression experiment set-up. (a) Square acrylic chamber and plunger used as a trough for compression experiments. The end of the plunger shaft away from the chamber is attached to a syringe pump pusher block. (b) Inset shows the $\sim\,1\,\text{mm}$ gap between the chamber and the plunger to allow for liquids to escape from the front. (c) Bi-axial compression of rafts is achieved in a funnel by draining the liquid.}
\label{fig:compression}
\end{figure*}

\textbf{Compression experiments}: Two different experimental setups are used for observing raft dynamics in compression - a square trough and a funnel. The square trough is an open-top acrylic chamber (Amazon) that compresses a raft uniaxially (Fig.\,\ref{fig:compression}a). The transparent and flat walls of the chamber allow for a detailed view of rafts and individual particles. Compression is achieved with a moving plate attached to the pusher block of a syringe pump (KD Scientific, Model: LEGATO110) with the help of a shaft. For all the compression experiments in the trough, a flow rate of 3\,mL/min is used, which translates to a linear speed of 55\,$\mu$m/s of the moving barrier. The compression direction is shown with an arrow in Fig.\,\ref{fig:compression}a. The width of the compressing plate is slightly smaller ($\approx\,0.5$\,mm on each side) than the internal width of the acrylic chamber, as shown in Fig.\,\ref{fig:compression}b. This ensures that when the raft is compressed, the fluid level stays at the same level on either side of the plunger. In the experiments with smaller particles (90\,$\mu$m-350\,$\mu$m), few particles also escape from this gap but do not affect the overall compression of the system. The funnel set-up is similar to \cite{Druecke2023} and shown in Fig.\,\ref{fig:compression}c. The compression of rafts is achieved by draining the fluid at a specific rate ($\dot{Q}\,=\,50\,\text{mL/min}$) with the help of a syringe pump. 

Buckling failure is preceded by the formation of folds in specific locations along the raft. Although we observed that the folds usually form at the funnel boundary or in the vicinity of the compressing barrier in the trough, folds can also grow near the raft center or at the other extremity of the trough (Movie S3 \cite{supp}). A similar observation of buckling near the moving boundary is explained as being equivalent to the well-known Janssen effect for dry granular media \cite{Cicuta2009}. A detailed analysis by Jambon-Puillet \cite{JambonPuillet2016a} comparing different raft areas and trough geometries showed that in the uniaxial compression experiments, the fold location is determined by the effective raft area relative to the trough width and the dimensionless parameter $\mathcal{M}$, which denotes the effective raft weight. Despite these reports, predicting the exact location of fold formation in the compression of a raft remains an open question. 

\linespread{1}

\textbf{Static equilibrium measurements}: For the static equilibrium position of particles at different fluid-fluid interfaces, a rectangular acrylic chamber with dimensions 6.5\,cm X 3\,cm X 2.5\,cm is used. Following our protocols for raft preparation, for particles smaller than $d\,<\,500\,\mu$m, lighter fluids are added before the particles are placed at the interface. Along with the order of fluid addition, the release height of particles is also an important factor that dictates the final equilibrium position of the particles. An in-house vacuum-controlled tweezer is used for this purpose. The tweezer is a syringe barrel with a hollow needle at the end connected to the vacuum pump via a flexible tube. Once the pump is turned on, the tube, barrel, and hollow needle stay in suction mode. Using an appropriate needle gauge (15 GA or 21 GA) according to the size of the particle, it is possible to pick up one particle with this mechanism. The distance between the needle holding the particle and the experimental chamber is adjusted to $\lesssim 6$\,mm via an adjustable steel platform on which the chamber is placed. A camera with a macro lens (Canon Mark IV fitted with a 65\,mm 1-5X zoom) is used for imaging. Turning the pump off and opening the pump valve releases the particle from the needle tip to the interface. Once the particle falls at the interface, we allow $\sim 1$\,min for the system to equilibrate before an image is taken. Three different particles were used for a reliable mean for each fluid-fluid interface and particle size. \\

\begin{figure*}[!ht]
\centering
\includegraphics[width=0.7\textwidth]{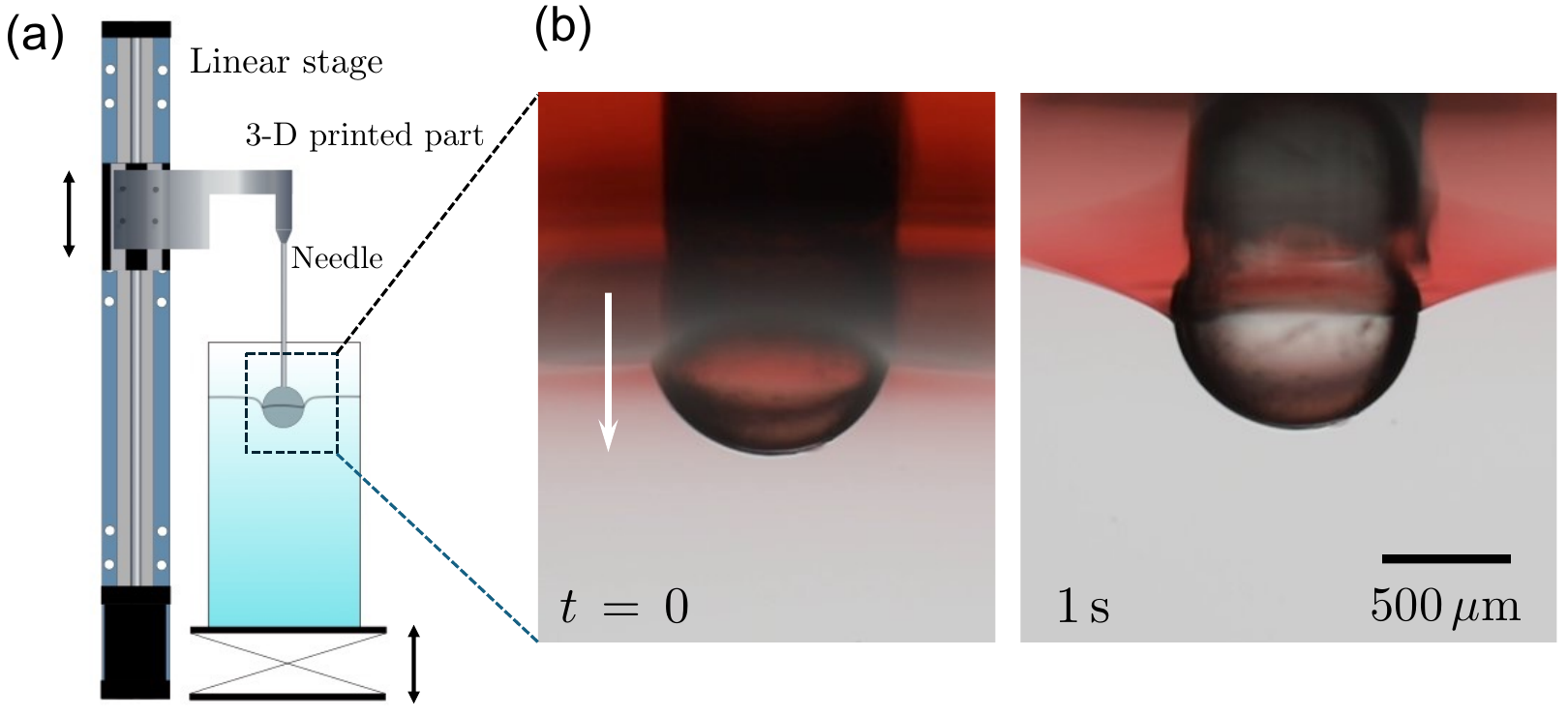}
\caption{Particle removal experiments. (a) Dynamic particle removal experiment set-up with a needle assembly attached to the sliding panel of a vertical linear stage. The particle is fixed to the end of the needle. (b) The equatorial plane of the particle is visible at the hexane-water interface in the $t=0$ frame. The white arrow shows the direction of the needle-particle assembly movement. The contact line position and interface deformation are visible after 1\,s.}
\label{fig:dynamic}
\end{figure*}

\textbf{Dynamic particle removal experiment}: The schematic of the whole set-up is shown in Fig.\,\ref{fig:dynamic}a. Our goal for this experiential set-up is to observe the contact line dynamics over the particle surface as a particle leaves the interface. The fluid-fluid interface is first prepared in the same rectangular chamber as before. $d\,>\,900\,\mu$m particles are used for better visualization. One such particle is glued (glass glue or super glue, Loctite) to the end of a needle (22G or 21G dispensing needle, Dispense All). The other end of the needle is locked in a custom 3D-printed fixture. This needle fixture is screwed to the sliding panel of a linear stage (FSL30, 100\,mm stroke length, Fuyu). The linear stage speed and direction are adjusted with a digital stepper driver (DM320T, StepperOnline) and a stepper motor drive controller (15-160\,VDC or 5-12\,VDC, Sywan). For our experiments, the lowest speed setting was used, roughly translating to a $750\,\mu\text{m}/s$ linear speed of the needle assembly. Videos of particles piercing the interface are recorded at 120\,fps with the help of a macro lens (Canon Mark IV fitted with a 65\,mm 1-5X zoom). Timestamped images from a representative experiment at a hexane-water interface with a $\sim\,900\,\mu$m particle are shown in Fig.\,\ref{fig:dynamic}b,c. \\

\linespread{1}

\begin{figure*}[!ht]
\centering
\includegraphics[width=0.55\textwidth]{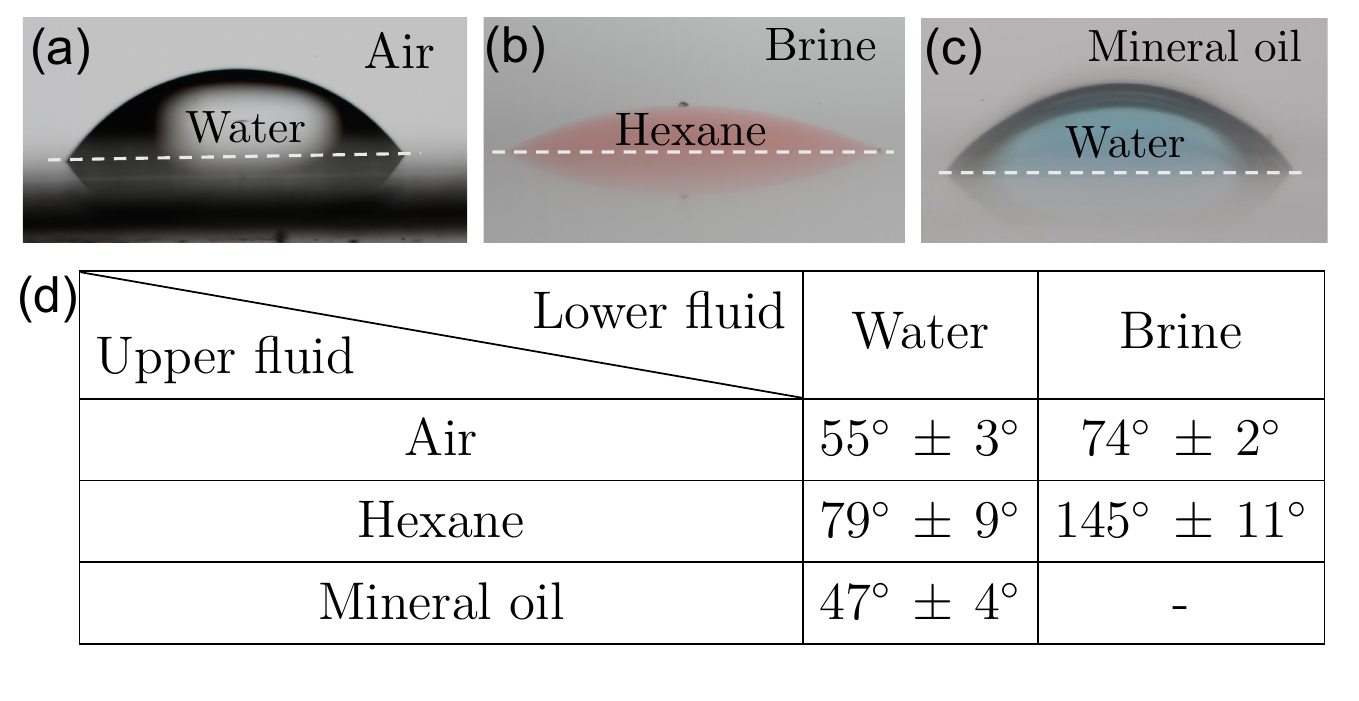}
\caption{Wettability experiments. (a--c,) The contact angle of different fluids on glass in varying fluid environments. The dashed white line denotes the position of the glass surface in all the images. (d) Table of contact angle values as measured from the experiments. The values are the average and standard deviation of three different measurements in each case.}
\label{fig:angles}
\end{figure*}

\textbf{Wettability measurements}: Due to the variability of our particle surfaces, we do not use a constant material contact angle value for defining particle equilibrium positions in our model. However, to calculate the interfacial energy contribution in the raft energy analysis (Eq.\,\ref{eq:raft}), a material contact angle value, $\theta_0$ is used for water on a glass surface in different fluids such as air, hexane, and mineral oil. For this, clean glass coverslips are selected because of their similar composition to our glass particles. Measuring the contact angle of water/brine on a glass slide in the air (Fig.\,\ref{fig:angles}a) or a mineral oil environment (Fig.\,\ref{fig:angles}c) is straightforward as water is the heavier fluid. The contact angle of water/brine on a glass surface in hexane is measured by placing upside-down glass slides wetted by hexane in a chamber filled with water/brine. When a glass slide wetted by hexane is placed in water, dewetting generates $<\,2\,\mu$L droplets of hexane as shown in Fig.\,\ref{fig:angles}b. We have tabulated the measurements in Fig.\,\ref{fig:angles}d. All the angles are measured from the heavier fluid to the lighter fluid.

\linespread{1}

\section{1D continuum model}

We consider an array of spherical particles and introduce the intrinsic coordinates $(s,\alpha)$ where $s$ is the arclength and $\alpha$ is the angle between the local tangent to the array and the horizontal axis, $x$. The position of the centerline of the array can then be parametrized as $\left[x(s), \eta(s)\right]$. The schematic for this setup is shown in Fig.\,\ref{fig:static}f and the inset Fig.\,\ref{fig:static}g. The total energy of the raft has contributions from the potential energy of the particles $U_p$, potential energy of the fluids $U_f$, and interfacial energy between two fluids $U_i$. In the above coordinate system, these energies will have the following forms:

\begin{subequations}\label{eq:raft} 
\begin{align}
    U_p &= (\rho_p-\rho_2) g (1-\beta)\forall_p \phi_s 
    \int_0^L (\eta + CG_2) \, ds \notag \\
    &\quad + (\rho_p-\rho_1) g \beta \forall_p \phi_s 
    \int_0^L (\eta - CG_1) \, ds, \\
    U_f &= \frac{(\rho_1 - \rho_2) g d}{2} 
    \int_0^L \eta^2 \cos(\alpha) \, ds, \\
    U_i &= \frac{\pi d^2 \gamma \phi_s}{2} 
    \int_0^L \bigg[ 
    1 + \cos^2\theta_\text{eq} 
    + \frac{3d^2 (\partial_s \alpha)^2}{8} \sin^4\theta_\text{eq}
    \bigg] \, ds
\end{align}
\end{subequations}
where $\forall_p=\pi d^3/6$ is the volume of a single spherical particle, $\phi_s=1/d$ is the number of particles per unit length along $s$ under a jammed condition, and $CG_1$ and $CG_2$ are the distances of the particle centroid below and above the contact line, respectively.

Similar to our particle-scale analysis, we can non-dimensionalize the energy terms, albeit differently. Here, the lengths are rescaled by $l_e=(K/\Delta\rho g)^{1/4}$ and energies by $Kd/l_e$, where $K= 3 \gamma  d^3 \phi_s \sin^4\theta_\text{eq}/8$. The non-dimensional energy terms become:

\begin{subequations}\label{eq:raft_nonD}
\begin{align}
   U_p^*&= \mathcal{M} \int_0^{L^*} \eta^* \,ds^*+c_1, \\
   U_f^*&=\frac{1}{2}\int_0^{L^*} {\eta^*}^2 \cos\alpha \,ds^*, \\
   U_i^*&=\frac{1}{2}\int_0^{L^*} ({\partial_{s^*}\alpha})^2 \cos\alpha \,ds^* +c_2.
\end{align}
\end{subequations}
where $c_1$, and $c_2$ are the constants that do not vary as the raft deforms, and $\mathcal{M}$ is a dimensionless number given by 
\begin{equation}
    \mathcal{M}=\frac{\pi d^2\phi_s}{6 l_e}\left[\frac{\rho_p-\rho_2}{\rho_1-\rho_2}-\beta\right].
\end{equation}
$\mathcal{M}$ contains all the essential material properties, including the particle size, surface wettability, and their position at the interface, along with fluid densities and interfacial tension. 

To find the equilibrium shape of the compressed particle array, we minimize the total nondimensional energy $U^*=U_p^*+U_f^*+U_i^*$ under two constraints:
\begin{equation}
    \Delta^*=\int_0^{L^*} P (1-\cos\alpha)ds^*, \quad \partial_{s^*}y^*=\sin\alpha,
\end{equation}
where $\Delta^*$ is the constraint of an imposed compression with an associated Lagrange multiplier $P$m and the second is a local constraint due to the use of intrinsic coordinates. With free-floating boundary conditions \cite{JambonPuillet2017} $ y^*(0)=y^*(L^*)=-\mathcal{M}, \quad \alpha(0)=\alpha(L^*)=0$, following the energy minimization procedure described in Ref.\,\cite{JambonPuillet2016} we obtain a single differential equation for the intrinsic angle $\alpha(s)$:
\begin{equation}
\begin{split}
    \partial^4_{s^*}\alpha+ \partial^2_{s^*}\alpha\left[\frac{3}{2}( \partial^2_{s^*}\alpha)^2-\frac{1}{2}( \partial^2_{s^*}\alpha(0))^2+P-\mathcal{M}\eta^*\right] \\
    +\sin\alpha \left(1-2\mathcal{M} \partial_{s^*}\alpha\right)=0.
\end{split}
\label{eq:shape_alpha}
\end{equation}
We solve the system of equations numerically using a continuation algorithm to follow the solutions as the parameters are varied \cite{JambonPuillet2016, JambonPuillet2017}. 

\section{Work required for expelling a single particle}

Here, we consider a spherical particle, so that $F_g=-\pi d^3\Delta\rho g/6$, $F_r=\pi d\gamma \sin\psi \sin\phi$, and $F_b=\pi d^3\Delta\rho g(2h\sin^2\psi/d+2/3-\cos\psi+\cos^3\psi/3)/8$. We note that $\psi$ and $\phi$ are the angular positions of the contact line relative to the center of the particle and the inclination
of the interface to the horizontal, respectively. We can use the geometrical relationship  $\phi=\theta+\psi-\pi$ to replace $\phi$ with contact angle $\theta$ \cite{Vella2006}.

Next, all the characteristic lengths are non-dimensionalized with the capillary length $l_c$, where $l_c=\sqrt{\gamma/\Delta\rho g}$, and the forces with $\Delta\rho gl_c^3$. From here onwards, the dimensionless parameters will be denoted with an asterisk. With that, the expression of the net force on a floating particle can be presented as:

\begin{equation}
\begin{split}
    F^*=d^*\pi^*\sin\psi \\
    +\frac{\pi{d^*}^3}{8}\left(\frac{2h_\infty^*}{d^*}\sin^2\psi+\frac{2}{3}-\cos\psi+\frac{1}{3}\cos^3\psi\right)
    -\frac{\pi}{6}D{d^*}^3,
\end{split}
\end{equation}
where $D=(\rho_p-\rho_2)/(\rho_1-\rho_2)$. The meniscus thickness $h^*$ can be obtained by solving the Laplace-Young equation: 
\begin{equation}
    h^*(r^*)=\frac{1}{r^*}\left(\frac{r^*h^*_{r^*}}{\left(1+{{h^*}_{r^*}}^2\right)^{1/2}}\right)_{r^*}
\end{equation}
with the following boundary conditions:
\begin{equation}
    h^*_{r^*}(r_0^*\sin\psi)=0, \, h^*(\infty)=0.
\end{equation}
Note that $r^*$ is the radial distance from the center of the particle, the lower script of $r^*$ denotes differentiation with respect to $r^*$, and $r_0^*=d^*/2$. The meniscus profile can be solved numerically by using MATLAB routine bvp4c and $h^*(r_0^*\sin\psi)\equiv h^*_\infty$, while $h^*_\infty$ varies with $\psi$ and $\theta$ \cite{vella2006load}.

\begin{figure}[ht!]  
\includegraphics[width=0.5\textwidth]{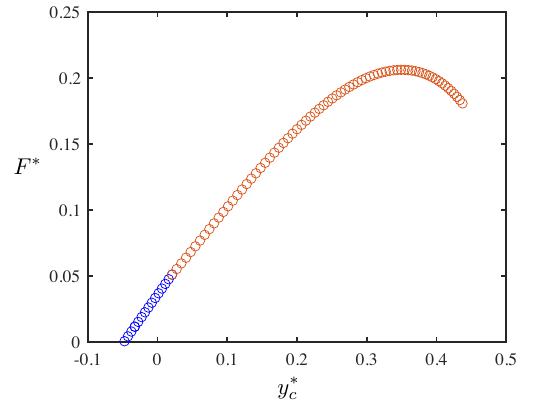}
\caption{Theoretical prediction of $F^*$ varies with different $y_c^*$ for water-hexane liquid combination and $d=900 \, \mu {\rm m}$. The blue and red circles are labeled as the first and second stages of the particle-expulsion process.}
\label{fig:Ftoyc}
\end{figure}

The work $\Delta E_p^*$ required to expel the particle off the liquid surface can be obtained by integrating the force $F^*$ along the vertical direction. The vertical displacement $y^*$ starts from the equilibrium position until the particle satisfies the expel condition. Here, we separate the particle expulsion process into two stages: (i) the contact angle increases from the equilibrium contact angle $\theta_0$ to the advancing contact angle $\theta_a$, while $\psi_0$ remains fixed, and (ii) once the contact angle reaches $\theta_a$, the contact line moves from $\psi_0$ until it detaches at the position $\psi_d$. Hence,
\begin{equation}
E_p^*=\int_{y_c^*(\theta_0)}^{y_c^*(\theta_a)} \,F^* \,{\rm d} y^* \bigg|_{\psi=\psi_0}+\int_{y_c^*(\psi_{a})}^{y_c^*(\psi_d)} F^* \,{\rm d} y^* \bigg|_{\theta=\theta_a}
\label{eq:particle_energy}
\end{equation}
where $y_c^*(\theta,\psi)=h^*_\infty(\theta,\psi)-(d^*)\cos{\psi}/2$ is the instantaneous position of the particle center; $\psi_d$ is the contact line position upon detachment, which is set when $F^*$ reaches a maximum value as shown in Fig.\,\ref{fig:Ftoyc}.

\begin{figure*}[ht!]  
\includegraphics[width=0.8\textwidth]{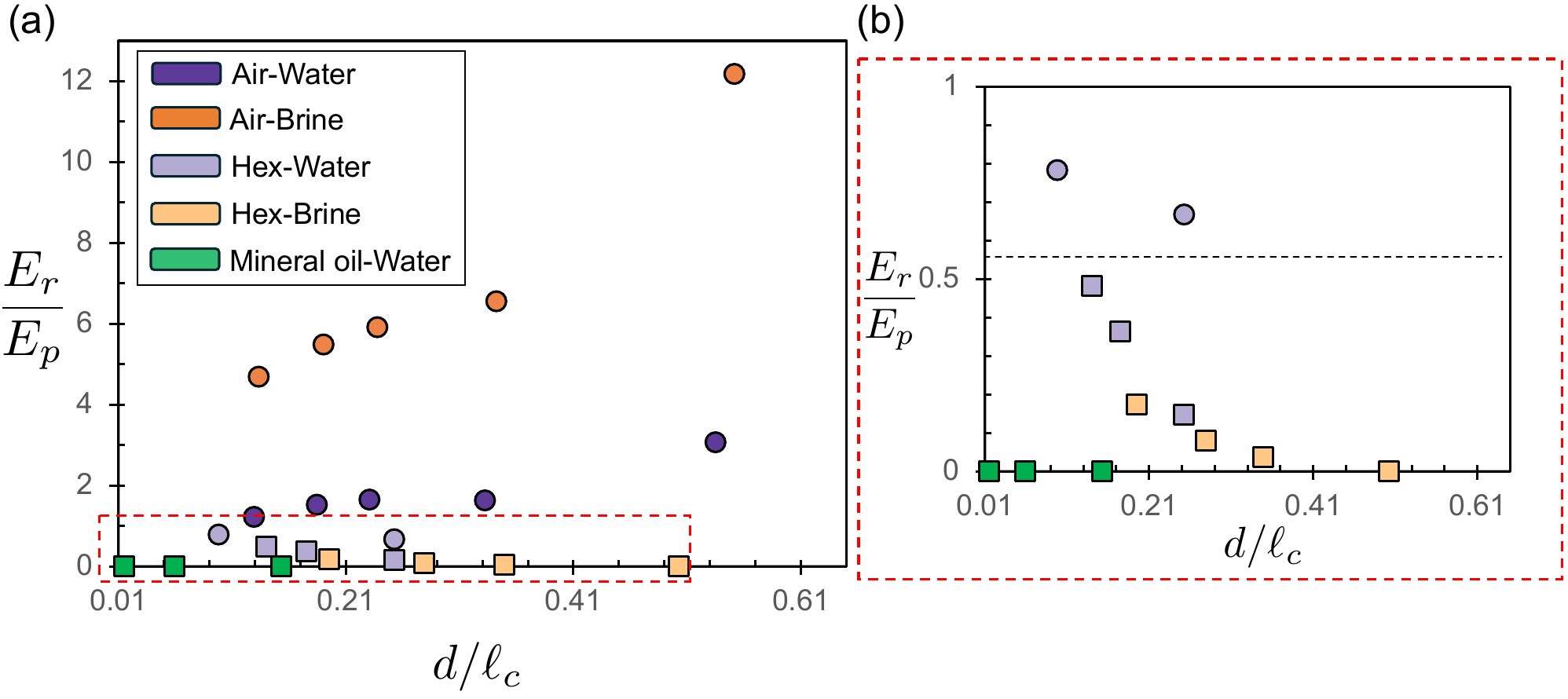}
\caption{Phase map of raft dynamics. (a) Reconstruction of the phase map in Fig.\,3c with an inverse energy ratio, $E_r/E_p$. Experimental observations for mineral oil-water (green) and larger particle data for the hexane-brine interface are shown here. (b) The transition point between the two modes is shown with the dashed horizontal line.}
\label{fig:phase-map}
\end{figure*}

\bibliography{Raft}

\begin{thebibliography}{51}%
\makeatletter
\providecommand \@ifxundefined [1]{%
 \@ifx{#1\undefined}
}%
\providecommand \@ifnum [1]{%
 \ifnum #1\expandafter \@firstoftwo
 \else \expandafter \@secondoftwo
 \fi
}%
\providecommand \@ifx [1]{%
 \ifx #1\expandafter \@firstoftwo
 \else \expandafter \@secondoftwo
 \fi
}%
\providecommand \natexlab [1]{#1}%
\providecommand \enquote  [1]{``#1''}%
\providecommand \bibnamefont  [1]{#1}%
\providecommand \bibfnamefont [1]{#1}%
\providecommand \citenamefont [1]{#1}%
\providecommand \href@noop [0]{\@secondoftwo}%
\providecommand \href [0]{\begingroup \@sanitize@url \@href}%
\providecommand \@href[1]{\@@startlink{#1}\@@href}%
\providecommand \@@href[1]{\endgroup#1\@@endlink}%
\providecommand \@sanitize@url [0]{\catcode `\\12\catcode `\$12\catcode `\&12\catcode `\#12\catcode `\^12\catcode `\_12\catcode `\%12\relax}%
\providecommand \@@startlink[1]{}%
\providecommand \@@endlink[0]{}%
\providecommand \url  [0]{\begingroup\@sanitize@url \@url }%
\providecommand \@url [1]{\endgroup\@href {#1}{\urlprefix }}%
\providecommand \urlprefix  [0]{URL }%
\providecommand \Eprint [0]{\href }%
\providecommand \doibase [0]{https://doi.org/}%
\providecommand \selectlanguage [0]{\@gobble}%
\providecommand \bibinfo  [0]{\@secondoftwo}%
\providecommand \bibfield  [0]{\@secondoftwo}%
\providecommand \translation [1]{[#1]}%
\providecommand \BibitemOpen [0]{}%
\providecommand \bibitemStop [0]{}%
\providecommand \bibitemNoStop [0]{.\EOS\space}%
\providecommand \EOS [0]{\spacefactor3000\relax}%
\providecommand \BibitemShut  [1]{\csname bibitem#1\endcsname}%
\let\auto@bib@innerbib\@empty
\bibitem [{\citenamefont {Nilsson}\ and\ \citenamefont {Pettersson}(2015)}]{Pettersson2015}%
  \BibitemOpen
  \bibfield  {author} {\bibinfo {author} {\bibfnamefont {A.}~\bibnamefont {Nilsson}}\ and\ \bibinfo {author} {\bibfnamefont {L.}~\bibnamefont {Pettersson}},\ }\bibfield  {title} {\bibinfo {title} {The structural origin of anomalous properties of liquid water},\ }\href@noop {} {\bibfield  {journal} {\bibinfo  {journal} {Nat. Commun.}\ }\textbf {\bibinfo {volume} {6}},\ \bibinfo {pages} {8998} (\bibinfo {year} {2015})}\BibitemShut {NoStop}%
\bibitem [{\citenamefont {Peskin}(1975)}]{Peskin1975}%
  \BibitemOpen
  \bibfield  {author} {\bibinfo {author} {\bibfnamefont {C.~S.}\ \bibnamefont {Peskin}},\ }\href@noop {} {\emph {\bibinfo {title} {Mathematical Aspects of Heart Physiology}}}\ (\bibinfo  {publisher} {Courant Institute of Mathematical Science Publication, New York},\ \bibinfo {year} {1975})\ Chap.\ \bibinfo {chapter} {VII}, pp.\ \bibinfo {pages} {268--278}\BibitemShut {NoStop}%
\bibitem [{\citenamefont {{Gal}}\ and\ \citenamefont {{Kronauer}}(2022)}]{Kronauer2022}%
  \BibitemOpen
  \bibfield  {author} {\bibinfo {author} {\bibfnamefont {A.}~\bibnamefont {{Gal}}}\ and\ \bibinfo {author} {\bibfnamefont {D.~J.}\ \bibnamefont {{Kronauer}}},\ }\bibfield  {title} {\bibinfo {title} {The emergence of a collective sensory response threshold in ant colonies},\ }\href@noop {} {\bibfield  {journal} {\bibinfo  {journal} {Proc. Natl. Acad. Sci. U.S.A.}\ }\textbf {\bibinfo {volume} {119}},\ \bibinfo {pages} {e2123076119} (\bibinfo {year} {2022})}\BibitemShut {NoStop}%
\bibitem [{\citenamefont {Peng}\ \emph {et~al.}(2021)\citenamefont {Peng}, \citenamefont {Liu},\ and\ \citenamefont {Cheng}}]{Peng2021_collective}%
  \BibitemOpen
  \bibfield  {author} {\bibinfo {author} {\bibfnamefont {Y.}~\bibnamefont {Peng}}, \bibinfo {author} {\bibfnamefont {Z.}~\bibnamefont {Liu}},\ and\ \bibinfo {author} {\bibfnamefont {X.}~\bibnamefont {Cheng}},\ }\bibfield  {title} {\bibinfo {title} {Imaging the emergence of bacterial turbulence: Phase diagram and transition kinetics},\ }\href@noop {} {\bibfield  {journal} {\bibinfo  {journal} {Sci. Adv.}\ }\textbf {\bibinfo {volume} {7}},\ \bibinfo {pages} {eabd1240} (\bibinfo {year} {2021})}\BibitemShut {NoStop}%
\bibitem [{\citenamefont {Mataric}(1993)}]{Mataric1993}%
  \BibitemOpen
  \bibfield  {author} {\bibinfo {author} {\bibfnamefont {M.~J.}\ \bibnamefont {Mataric}},\ }\bibfield  {title} {\bibinfo {title} {Designing emergent behaviors: From local interactions to collective intelligence},\ }in\ \href@noop {} {\emph {\bibinfo {booktitle} {From Animals to Animats 2: Proceedings of the Second International Conference on Simulation of Adaptive Behavior}}},\ \bibinfo {editor} {edited by\ \bibinfo {editor} {\bibfnamefont {J.-A.}\ \bibnamefont {Meyer}}, \bibinfo {editor} {\bibfnamefont {H.~L.}\ \bibnamefont {Roitblat}},\ and\ \bibinfo {editor} {\bibfnamefont {S.~W.}\ \bibnamefont {Wilson}}}\ (\bibinfo  {publisher} {The MIT Press},\ \bibinfo {address} {Cambridge, MA},\ \bibinfo {year} {1993})\ pp.\ \bibinfo {pages} {432--441}\BibitemShut {NoStop}%
\bibitem [{\citenamefont {Kadic}\ \emph {et~al.}(2019)\citenamefont {Kadic}, \citenamefont {Milton}, \citenamefont {van Hecke},\ and\ \citenamefont {Wegener}}]{Milton2019}%
  \BibitemOpen
  \bibfield  {author} {\bibinfo {author} {\bibfnamefont {M.}~\bibnamefont {Kadic}}, \bibinfo {author} {\bibfnamefont {G.~W.}\ \bibnamefont {Milton}}, \bibinfo {author} {\bibfnamefont {M.}~\bibnamefont {van Hecke}},\ and\ \bibinfo {author} {\bibfnamefont {M.}~\bibnamefont {Wegener}},\ }\bibfield  {title} {\bibinfo {title} {3d metamaterials},\ }\href@noop {} {\bibfield  {journal} {\bibinfo  {journal} {Nat. Rev. Phys.}\ }\textbf {\bibinfo {volume} {1}},\ \bibinfo {pages} {198} (\bibinfo {year} {2019})}\BibitemShut {NoStop}%
\bibitem [{\citenamefont {{Vella}}\ and\ \citenamefont {{Mahadevan}}(2005)}]{Vella2005}%
  \BibitemOpen
  \bibfield  {author} {\bibinfo {author} {\bibfnamefont {D.}~\bibnamefont {{Vella}}}\ and\ \bibinfo {author} {\bibfnamefont {L.}~\bibnamefont {{Mahadevan}}},\ }\bibfield  {title} {\bibinfo {title} {{The ``Cheerios effect''}},\ }\href@noop {} {\bibfield  {journal} {\bibinfo  {journal} {Am. J. Phys.}\ }\textbf {\bibinfo {volume} {73}},\ \bibinfo {pages} {817} (\bibinfo {year} {2005})}\BibitemShut {NoStop}%
\bibitem [{\citenamefont {Vella}\ \emph {et~al.}(2004)\citenamefont {Vella}, \citenamefont {Aussillous},\ and\ \citenamefont {Mahadevan}}]{Vella2004a}%
  \BibitemOpen
  \bibfield  {author} {\bibinfo {author} {\bibfnamefont {D.}~\bibnamefont {Vella}}, \bibinfo {author} {\bibfnamefont {P.}~\bibnamefont {Aussillous}},\ and\ \bibinfo {author} {\bibfnamefont {L.}~\bibnamefont {Mahadevan}},\ }\bibfield  {title} {\bibinfo {title} {Elasticity of an interfacial particle raft},\ }\href@noop {} {\bibfield  {journal} {\bibinfo  {journal} {Europhys. Lett.}\ }\textbf {\bibinfo {volume} {68}},\ \bibinfo {pages} {212} (\bibinfo {year} {2004})}\BibitemShut {NoStop}%
\bibitem [{\citenamefont {Monteux}\ \emph {et~al.}(2007)\citenamefont {Monteux}, \citenamefont {Kirkwood}, \citenamefont {Xu}, \citenamefont {Jung},\ and\ \citenamefont {Fuller}}]{Monteux2007}%
  \BibitemOpen
  \bibfield  {author} {\bibinfo {author} {\bibfnamefont {C.}~\bibnamefont {Monteux}}, \bibinfo {author} {\bibfnamefont {J.}~\bibnamefont {Kirkwood}}, \bibinfo {author} {\bibfnamefont {H.}~\bibnamefont {Xu}}, \bibinfo {author} {\bibfnamefont {E.}~\bibnamefont {Jung}},\ and\ \bibinfo {author} {\bibfnamefont {G.~G.}\ \bibnamefont {Fuller}},\ }\bibfield  {title} {\bibinfo {title} {Determining the mechanical response of particle-laden fluid interfaces using surface pressure isotherms and bulk pressure measurements of droplets},\ }\href {https://doi.org/10.1039/B708962G} {\bibfield  {journal} {\bibinfo  {journal} {Phys. Chem. Chem. Phys.}\ }\textbf {\bibinfo {volume} {9}},\ \bibinfo {pages} {6344} (\bibinfo {year} {2007})}\BibitemShut {NoStop}%
\bibitem [{\citenamefont {Leahy}\ \emph {et~al.}(2010)\citenamefont {Leahy}, \citenamefont {Pocivavsek}, \citenamefont {Meron}, \citenamefont {Lam}, \citenamefont {Salas}, \citenamefont {Viccaro}, \citenamefont {Lee},\ and\ \citenamefont {Lin}}]{Leahy2010a}%
  \BibitemOpen
  \bibfield  {author} {\bibinfo {author} {\bibfnamefont {B.~D.}\ \bibnamefont {Leahy}}, \bibinfo {author} {\bibfnamefont {L.}~\bibnamefont {Pocivavsek}}, \bibinfo {author} {\bibfnamefont {M.}~\bibnamefont {Meron}}, \bibinfo {author} {\bibfnamefont {K.~L.}\ \bibnamefont {Lam}}, \bibinfo {author} {\bibfnamefont {D.}~\bibnamefont {Salas}}, \bibinfo {author} {\bibfnamefont {P.~J.}\ \bibnamefont {Viccaro}}, \bibinfo {author} {\bibfnamefont {K.~Y.~C.}\ \bibnamefont {Lee}},\ and\ \bibinfo {author} {\bibfnamefont {B.~H.}\ \bibnamefont {Lin}},\ }\bibfield  {title} {\bibinfo {title} {Geometric stability and elastic response of a supported nanoparticle film},\ }\href@noop {} {\bibfield  {journal} {\bibinfo  {journal} {Phys. Rev. Lett.}\ }\textbf {\bibinfo {volume} {105}},\ \bibinfo {pages} {4} (\bibinfo {year} {2010})}\BibitemShut {NoStop}%
\bibitem [{\citenamefont {{Proti{\`e}re}}\ \emph {et~al.}(2017)\citenamefont {{Proti{\`e}re}}, \citenamefont {{Josserand}}, \citenamefont {{Aristoff}}, \citenamefont {{Stone}},\ and\ \citenamefont {{Abkarian}}}]{Protiere2017}%
  \BibitemOpen
  \bibfield  {author} {\bibinfo {author} {\bibfnamefont {S.}~\bibnamefont {{Proti{\`e}re}}}, \bibinfo {author} {\bibfnamefont {C.}~\bibnamefont {{Josserand}}}, \bibinfo {author} {\bibfnamefont {J.~M.}\ \bibnamefont {{Aristoff}}}, \bibinfo {author} {\bibfnamefont {H.~A.}\ \bibnamefont {{Stone}}},\ and\ \bibinfo {author} {\bibfnamefont {M.}~\bibnamefont {{Abkarian}}},\ }\bibfield  {title} {\bibinfo {title} {{Sinking a granular raft}},\ }\href {https://doi.org/10.1103/PhysRevLett.118.108001} {\bibfield  {journal} {\bibinfo  {journal} {Phys. Rev. Lett.}\ }\textbf {\bibinfo {volume} {118}},\ \bibinfo {eid} {108001} (\bibinfo {year} {2017})}\BibitemShut {NoStop}%
\bibitem [{\citenamefont {Druecke}\ \emph {et~al.}(2023)\citenamefont {Druecke}, \citenamefont {Mukherjee}, \citenamefont {Cheng},\ and\ \citenamefont {Lee}}]{Druecke2023}%
  \BibitemOpen
  \bibfield  {author} {\bibinfo {author} {\bibfnamefont {B.~C.}\ \bibnamefont {Druecke}}, \bibinfo {author} {\bibfnamefont {R.}~\bibnamefont {Mukherjee}}, \bibinfo {author} {\bibfnamefont {X.}~\bibnamefont {Cheng}},\ and\ \bibinfo {author} {\bibfnamefont {S.}~\bibnamefont {Lee}},\ }\bibfield  {title} {\bibinfo {title} {Collapse of a granular raft: transition from single particle falling to collective creasing},\ }\href@noop {} {\bibfield  {journal} {\bibinfo  {journal} {Phys. Rev. Fluids}\ }\textbf {\bibinfo {volume} {8}},\ \bibinfo {pages} {024003} (\bibinfo {year} {2023})}\BibitemShut {NoStop}%
\bibitem [{sup()}]{supp}%
  \BibitemOpen
  \href@noop {} {}\bibinfo {note} {See Supplemental Material at URL-to-be-inserted for Movies S1–S5 and a description of the movies.}\BibitemShut {Stop}%
\bibitem [{\citenamefont {{Huang}}\ \emph {et~al.}(2007)\citenamefont {{Huang}}, \citenamefont {{Juszkiewicz}}, \citenamefont {{de Jeu}}, \citenamefont {{Cerda}}, \citenamefont {{Emrick}},\ and\ \citenamefont {{Menon}}}]{Huang2007}%
  \BibitemOpen
  \bibfield  {author} {\bibinfo {author} {\bibfnamefont {J.}~\bibnamefont {{Huang}}}, \bibinfo {author} {\bibfnamefont {M.}~\bibnamefont {{Juszkiewicz}}}, \bibinfo {author} {\bibfnamefont {W.~H.}\ \bibnamefont {{de Jeu}}}, \bibinfo {author} {\bibfnamefont {E.}~\bibnamefont {{Cerda}}}, \bibinfo {author} {\bibfnamefont {T.}~\bibnamefont {{Emrick}}},\ and\ \bibinfo {author} {\bibfnamefont {N.}~\bibnamefont {{Menon}}},\ }\bibfield  {title} {\bibinfo {title} {{Capillary wrinkling of floating thin polymer films}},\ }\href {https://doi.org/10.1126/science.1144616} {\bibfield  {journal} {\bibinfo  {journal} {Science}\ }\textbf {\bibinfo {volume} {317}},\ \bibinfo {pages} {650} (\bibinfo {year} {2007})}\BibitemShut {NoStop}%
\bibitem [{\citenamefont {{Pocivavsek}}\ \emph {et~al.}(2008)\citenamefont {{Pocivavsek}}, \citenamefont {{Dellsy}}, \citenamefont {{Kern}}, \citenamefont {{Johnson}}, \citenamefont {{Lin}}, \citenamefont {{Lee}},\ and\ \citenamefont {{Cerda}}}]{Pocivavsek2008a}%
  \BibitemOpen
  \bibfield  {author} {\bibinfo {author} {\bibfnamefont {L.}~\bibnamefont {{Pocivavsek}}}, \bibinfo {author} {\bibfnamefont {R.}~\bibnamefont {{Dellsy}}}, \bibinfo {author} {\bibfnamefont {A.}~\bibnamefont {{Kern}}}, \bibinfo {author} {\bibfnamefont {S.}~\bibnamefont {{Johnson}}}, \bibinfo {author} {\bibfnamefont {B.}~\bibnamefont {{Lin}}}, \bibinfo {author} {\bibfnamefont {K.~Y.~C.}\ \bibnamefont {{Lee}}},\ and\ \bibinfo {author} {\bibfnamefont {E.}~\bibnamefont {{Cerda}}},\ }\bibfield  {title} {\bibinfo {title} {{Stress and fold localization in thin elastic membranes}},\ }\href {https://doi.org/10.1126/science.1154069} {\bibfield  {journal} {\bibinfo  {journal} {Science}\ }\textbf {\bibinfo {volume} {320}},\ \bibinfo {pages} {912} (\bibinfo {year} {2008})}\BibitemShut {NoStop}%
\bibitem [{\citenamefont {{Razavi}}\ and\ \citenamefont {{Wang}}(2015)}]{Razavi2015}%
  \BibitemOpen
  \bibfield  {author} {\bibinfo {author} {\bibfnamefont {M.~J.}\ \bibnamefont {{Razavi}}}\ and\ \bibinfo {author} {\bibfnamefont {X.}~\bibnamefont {{Wang}}},\ }\bibfield  {title} {\bibinfo {title} {{Morphological patterns of a growing biological tube in a confined environment with contacting boundary}},\ }\href@noop {} {\bibfield  {journal} {\bibinfo  {journal} {RSC Adv.}\ }\textbf {\bibinfo {volume} {5}},\ \bibinfo {pages} {7440} (\bibinfo {year} {2015})}\BibitemShut {NoStop}%
\bibitem [{\citenamefont {{Jambon-Puillet}}\ \emph {et~al.}(2017)\citenamefont {{Jambon-Puillet}}, \citenamefont {{Josserand}},\ and\ \citenamefont {{Proti{\`e}re}}}]{JambonPuillet2017}%
  \BibitemOpen
  \bibfield  {author} {\bibinfo {author} {\bibfnamefont {E.}~\bibnamefont {{Jambon-Puillet}}}, \bibinfo {author} {\bibfnamefont {C.}~\bibnamefont {{Josserand}}},\ and\ \bibinfo {author} {\bibfnamefont {S.}~\bibnamefont {{Proti{\`e}re}}},\ }\bibfield  {title} {\bibinfo {title} {{Wrinkles, folds, and plasticity in granular rafts}},\ }\href@noop {} {\bibfield  {journal} {\bibinfo  {journal} {Phys. Rev. Mater.}\ }\textbf {\bibinfo {volume} {1}},\ \bibinfo {pages} {042601} (\bibinfo {year} {2017})}\BibitemShut {NoStop}%
\bibitem [{\citenamefont {{Cicuta}}\ and\ \citenamefont {{Vella}}(2009)}]{Cicuta2009}%
  \BibitemOpen
  \bibfield  {author} {\bibinfo {author} {\bibfnamefont {P.}~\bibnamefont {{Cicuta}}}\ and\ \bibinfo {author} {\bibfnamefont {D.}~\bibnamefont {{Vella}}},\ }\bibfield  {title} {\bibinfo {title} {{Granular character of particle rafts}},\ }\href@noop {} {\bibfield  {journal} {\bibinfo  {journal} {Phys. Rev. Lett.}\ }\textbf {\bibinfo {volume} {102}},\ \bibinfo {pages} {138302} (\bibinfo {year} {2009})}\BibitemShut {NoStop}%
\bibitem [{\citenamefont {Binks}\ and\ \citenamefont {Horozov}(2005)}]{Binks2005}%
  \BibitemOpen
  \bibfield  {author} {\bibinfo {author} {\bibfnamefont {B.~P.}\ \bibnamefont {Binks}}\ and\ \bibinfo {author} {\bibfnamefont {T.~S.}\ \bibnamefont {Horozov}},\ }\bibfield  {title} {\bibinfo {title} {Aqueous foams stabilized solely by silica nanoparticles},\ }\href@noop {} {\bibfield  {journal} {\bibinfo  {journal} {Angew. Chem. Int. Ed.}\ }\textbf {\bibinfo {volume} {44}},\ \bibinfo {pages} {3722} (\bibinfo {year} {2005})}\BibitemShut {NoStop}%
\bibitem [{\citenamefont {{Abkarian}}\ \emph {et~al.}(2013)\citenamefont {{Abkarian}}, \citenamefont {{Proti{\`e}re}}, \citenamefont {{Aristoff}},\ and\ \citenamefont {{Stone}}}]{Abkarian2013}%
  \BibitemOpen
  \bibfield  {author} {\bibinfo {author} {\bibfnamefont {M.}~\bibnamefont {{Abkarian}}}, \bibinfo {author} {\bibfnamefont {S.}~\bibnamefont {{Proti{\`e}re}}}, \bibinfo {author} {\bibfnamefont {J.~M.}\ \bibnamefont {{Aristoff}}},\ and\ \bibinfo {author} {\bibfnamefont {H.~A.}\ \bibnamefont {{Stone}}},\ }\bibfield  {title} {\bibinfo {title} {{Gravity-induced encapsulation of liquids by destabilization of granular rafts}},\ }\href@noop {} {\bibfield  {journal} {\bibinfo  {journal} {Nat. Commun.}\ }\textbf {\bibinfo {volume} {4}},\ \bibinfo {pages} {1895} (\bibinfo {year} {2013})}\BibitemShut {NoStop}%
\bibitem [{\citenamefont {Pickering}(1907)}]{Pickering1907}%
  \BibitemOpen
  \bibfield  {author} {\bibinfo {author} {\bibfnamefont {S.~U.}\ \bibnamefont {Pickering}},\ }\bibfield  {title} {\bibinfo {title} {Emulsions},\ }\href@noop {} {\bibfield  {journal} {\bibinfo  {journal} {J. Chem. Soc.}\ }\textbf {\bibinfo {volume} {91}},\ \bibinfo {pages} {2001} (\bibinfo {year} {1907})}\BibitemShut {NoStop}%
\bibitem [{\citenamefont {Schutt}\ \emph {et~al.}(2003)\citenamefont {Schutt}, \citenamefont {Klein}, \citenamefont {Mattrey},\ and\ \citenamefont {Riess}}]{Reiss2003}%
  \BibitemOpen
  \bibfield  {author} {\bibinfo {author} {\bibfnamefont {E.~G.}\ \bibnamefont {Schutt}}, \bibinfo {author} {\bibfnamefont {D.~H.}\ \bibnamefont {Klein}}, \bibinfo {author} {\bibfnamefont {R.~M.}\ \bibnamefont {Mattrey}},\ and\ \bibinfo {author} {\bibfnamefont {J.~G.}\ \bibnamefont {Riess}},\ }\bibfield  {title} {\bibinfo {title} {{Injectable microbubbles as contrast agents for diagnostic ultrasound imaging: the key role of perfluorochemicals}},\ }\href@noop {} {\bibfield  {journal} {\bibinfo  {journal} {Angew. Chem. Int. Ed.}\ }\textbf {\bibinfo {volume} {42}},\ \bibinfo {pages} {3218} (\bibinfo {year} {2003})}\BibitemShut {NoStop}%
\bibitem [{\citenamefont {Stratford}\ \emph {et~al.}(2005)\citenamefont {Stratford}, \citenamefont {Adhikari}, \citenamefont {Pagonabarraga}, \citenamefont {Desplat},\ and\ \citenamefont {Cates}}]{Stratford2005_bijels}%
  \BibitemOpen
  \bibfield  {author} {\bibinfo {author} {\bibfnamefont {K.}~\bibnamefont {Stratford}}, \bibinfo {author} {\bibfnamefont {R.}~\bibnamefont {Adhikari}}, \bibinfo {author} {\bibfnamefont {I.}~\bibnamefont {Pagonabarraga}}, \bibinfo {author} {\bibfnamefont {J.-C.}\ \bibnamefont {Desplat}},\ and\ \bibinfo {author} {\bibfnamefont {M.~E.}\ \bibnamefont {Cates}},\ }\bibfield  {title} {\bibinfo {title} {Colloidal jamming at interfaces: a route to fluid-bicontinuous gels},\ }\href@noop {} {\bibfield  {journal} {\bibinfo  {journal} {Science}\ }\textbf {\bibinfo {volume} {309}},\ \bibinfo {pages} {2198} (\bibinfo {year} {2005})}\BibitemShut {NoStop}%
\bibitem [{\citenamefont {Cates}\ and\ \citenamefont {Clegg}(2008)}]{Cates2008_bijels}%
  \BibitemOpen
  \bibfield  {author} {\bibinfo {author} {\bibfnamefont {M.~E.}\ \bibnamefont {Cates}}\ and\ \bibinfo {author} {\bibfnamefont {P.~S.}\ \bibnamefont {Clegg}},\ }\bibfield  {title} {\bibinfo {title} {Bijels: a new class of soft materials},\ }\href@noop {} {\bibfield  {journal} {\bibinfo  {journal} {Soft Matter}\ }\textbf {\bibinfo {volume} {4}},\ \bibinfo {pages} {2132} (\bibinfo {year} {2008})}\BibitemShut {NoStop}%
\bibitem [{\citenamefont {Bai}\ \emph {et~al.}(2015)\citenamefont {Bai}, \citenamefont {Fruehwirth}, \citenamefont {Cheng},\ and\ \citenamefont {Macosko}}]{Bai2015_bijels}%
  \BibitemOpen
  \bibfield  {author} {\bibinfo {author} {\bibfnamefont {L.}~\bibnamefont {Bai}}, \bibinfo {author} {\bibfnamefont {J.~W.}\ \bibnamefont {Fruehwirth}}, \bibinfo {author} {\bibfnamefont {X.}~\bibnamefont {Cheng}},\ and\ \bibinfo {author} {\bibfnamefont {C.~W.}\ \bibnamefont {Macosko}},\ }\bibfield  {title} {\bibinfo {title} {Dynamics and rheology of nonpolar bijels},\ }\href@noop {} {\bibfield  {journal} {\bibinfo  {journal} {Soft Matter}\ }\textbf {\bibinfo {volume} {11}},\ \bibinfo {pages} {5282} (\bibinfo {year} {2015})}\BibitemShut {NoStop}%
\bibitem [{\citenamefont {Crossley}\ \emph {et~al.}(2010)\citenamefont {Crossley}, \citenamefont {Faria}, \citenamefont {Shen},\ and\ \citenamefont {Resasco}}]{Resasco2010}%
  \BibitemOpen
  \bibfield  {author} {\bibinfo {author} {\bibfnamefont {S.}~\bibnamefont {Crossley}}, \bibinfo {author} {\bibfnamefont {J.}~\bibnamefont {Faria}}, \bibinfo {author} {\bibfnamefont {M.}~\bibnamefont {Shen}},\ and\ \bibinfo {author} {\bibfnamefont {D.~E.}\ \bibnamefont {Resasco}},\ }\bibfield  {title} {\bibinfo {title} {Solid nanoparticles that catalyze biofuel upgrade reactions at the water/oil interface},\ }\href@noop {} {\bibfield  {journal} {\bibinfo  {journal} {Science}\ }\textbf {\bibinfo {volume} {227}},\ \bibinfo {pages} {68} (\bibinfo {year} {2010})}\BibitemShut {NoStop}%
\bibitem [{\citenamefont {Poulichet}\ and\ \citenamefont {Garbin}(2015)}]{Garbin2015}%
  \BibitemOpen
  \bibfield  {author} {\bibinfo {author} {\bibfnamefont {V.}~\bibnamefont {Poulichet}}\ and\ \bibinfo {author} {\bibfnamefont {V.}~\bibnamefont {Garbin}},\ }\bibfield  {title} {\bibinfo {title} {Ultrafast desorption of colloidal particles from fluid interfaces},\ }\href@noop {} {\bibfield  {journal} {\bibinfo  {journal} {Proc. Natl. Acad. Sci. U.S.A.}\ }\textbf {\bibinfo {volume} {112}},\ \bibinfo {pages} {5932} (\bibinfo {year} {2015})}\BibitemShut {NoStop}%
\bibitem [{\citenamefont {Bord\'{a}cs}\ \emph {et~al.}(2006)\citenamefont {Bord\'{a}cs}, \citenamefont {Agod},\ and\ \citenamefont {H\'{o}rv\"{o}lgyi}}]{Bordacs2006}%
  \BibitemOpen
  \bibfield  {author} {\bibinfo {author} {\bibfnamefont {S.}~\bibnamefont {Bord\'{a}cs}}, \bibinfo {author} {\bibfnamefont {A.}~\bibnamefont {Agod}},\ and\ \bibinfo {author} {\bibfnamefont {Z.}~\bibnamefont {H\'{o}rv\"{o}lgyi}},\ }\bibfield  {title} {\bibinfo {title} {Compression of langmuir films composed of fine particles: Collapse mechanism and wettability},\ }\href@noop {} {\bibfield  {journal} {\bibinfo  {journal} {Langmuir}\ }\textbf {\bibinfo {volume} {22}},\ \bibinfo {pages} {6944} (\bibinfo {year} {2006})}\BibitemShut {NoStop}%
\bibitem [{\citenamefont {{Planchette}}\ \emph {et~al.}(2012)\citenamefont {{Planchette}}, \citenamefont {{Lorenceau}},\ and\ \citenamefont {{Biance}}}]{Planchette2012}%
  \BibitemOpen
  \bibfield  {author} {\bibinfo {author} {\bibfnamefont {C.}~\bibnamefont {{Planchette}}}, \bibinfo {author} {\bibfnamefont {E.}~\bibnamefont {{Lorenceau}}},\ and\ \bibinfo {author} {\bibfnamefont {A.-L.}\ \bibnamefont {{Biance}}},\ }\bibfield  {title} {\bibinfo {title} {{Surface wave on a particle raft}},\ }\href {https://doi.org/10.1039/c2sm06859a} {\bibfield  {journal} {\bibinfo  {journal} {Soft Matter}\ }\textbf {\bibinfo {volume} {8}},\ \bibinfo {pages} {2444} (\bibinfo {year} {2012})}\BibitemShut {NoStop}%
\bibitem [{\citenamefont {{Gu}}\ and\ \citenamefont {{Botto}}(2018)}]{Gu2018}%
  \BibitemOpen
  \bibfield  {author} {\bibinfo {author} {\bibfnamefont {C.}~\bibnamefont {{Gu}}}\ and\ \bibinfo {author} {\bibfnamefont {L.}~\bibnamefont {{Botto}}},\ }\bibfield  {title} {\bibinfo {title} {{Buckling vs. particle desorption in a particle-covered drop subject to compressive surface stresses: a simulation study}},\ }\href {https://doi.org/10.1039/C7SM01912B} {\bibfield  {journal} {\bibinfo  {journal} {Soft Matter}\ }\textbf {\bibinfo {volume} {14}},\ \bibinfo {pages} {711} (\bibinfo {year} {2018})}\BibitemShut {NoStop}%
\bibitem [{\citenamefont {Majérus}\ \emph {et~al.}(2020)\citenamefont {Majérus}, \citenamefont {Lehuédé}, \citenamefont {Biron}, \citenamefont {Alloteau}, \citenamefont {Narayanasamy},\ and\ \citenamefont {Caurant}}]{Caurant2020}%
  \BibitemOpen
  \bibfield  {author} {\bibinfo {author} {\bibfnamefont {O.}~\bibnamefont {Majérus}}, \bibinfo {author} {\bibfnamefont {P.}~\bibnamefont {Lehuédé}}, \bibinfo {author} {\bibfnamefont {I.}~\bibnamefont {Biron}}, \bibinfo {author} {\bibfnamefont {F.}~\bibnamefont {Alloteau}}, \bibinfo {author} {\bibfnamefont {S.}~\bibnamefont {Narayanasamy}},\ and\ \bibinfo {author} {\bibfnamefont {D.}~\bibnamefont {Caurant}},\ }\bibfield  {title} {\bibinfo {title} {Glass alteration in atmospheric conditions: crossing perspectives from cultural heritage, glass industry, and nuclear waste management},\ }\href@noop {} {\bibfield  {journal} {\bibinfo  {journal} {npj Mater. Degrad.}\ }\textbf {\bibinfo {volume} {4}},\ \bibinfo {pages} {27} (\bibinfo {year} {2020})}\BibitemShut {NoStop}%
\bibitem [{\citenamefont {{Varshney}}\ \emph {et~al.}(2012)\citenamefont {{Varshney}}, \citenamefont {{Sane}}, \citenamefont {{Ghosh}},\ and\ \citenamefont {{Bhattacharya}}}]{Varshney2012}%
  \BibitemOpen
  \bibfield  {author} {\bibinfo {author} {\bibfnamefont {A.}~\bibnamefont {{Varshney}}}, \bibinfo {author} {\bibfnamefont {A.}~\bibnamefont {{Sane}}}, \bibinfo {author} {\bibfnamefont {S.}~\bibnamefont {{Ghosh}}},\ and\ \bibinfo {author} {\bibfnamefont {S.}~\bibnamefont {{Bhattacharya}}},\ }\bibfield  {title} {\bibinfo {title} {{Amorphous to amorphous transition in particle rafts}},\ }\href@noop {} {\bibfield  {journal} {\bibinfo  {journal} {Phys. Rev. E}\ }\textbf {\bibinfo {volume} {86}},\ \bibinfo {pages} {031402} (\bibinfo {year} {2012})}\BibitemShut {NoStop}%
\bibitem [{\citenamefont {{Lagubeau}}\ \emph {et~al.}(2014)\citenamefont {{Lagubeau}}, \citenamefont {{Rescaglio}},\ and\ \citenamefont {{Melo}}}]{Lagubeau2014}%
  \BibitemOpen
  \bibfield  {author} {\bibinfo {author} {\bibfnamefont {G.}~\bibnamefont {{Lagubeau}}}, \bibinfo {author} {\bibfnamefont {A.}~\bibnamefont {{Rescaglio}}},\ and\ \bibinfo {author} {\bibfnamefont {F.}~\bibnamefont {{Melo}}},\ }\bibfield  {title} {\bibinfo {title} {{Armoring a droplet: Soft jamming of a dense granular interface}},\ }\href {https://doi.org/10.1103/PhysRevE.90.030201} {\bibfield  {journal} {\bibinfo  {journal} {Phys. Rev. E}\ }\textbf {\bibinfo {volume} {90}},\ \bibinfo {eid} {030201} (\bibinfo {year} {2014})}\BibitemShut {NoStop}%
\bibitem [{\citenamefont {Jambon-Puillet}(2016)}]{JambonPuillet2016a}%
  \BibitemOpen
  \bibfield  {author} {\bibinfo {author} {\bibfnamefont {E.}~\bibnamefont {Jambon-Puillet}},\ }\emph {\bibinfo {title} {Folds in floating membranes : from elastic sheets to granular rafts}},\ \href@noop {} {Ph.D. thesis},\ \bibinfo  {school} {Université Pierre et Marie Curie - Paris} (\bibinfo {year} {2016})\BibitemShut {NoStop}%
\bibitem [{\citenamefont {{Cerda}}\ and\ \citenamefont {{Mahadevan}}(2003)}]{Cerda2003}%
  \BibitemOpen
  \bibfield  {author} {\bibinfo {author} {\bibfnamefont {E.}~\bibnamefont {{Cerda}}}\ and\ \bibinfo {author} {\bibfnamefont {L.}~\bibnamefont {{Mahadevan}}},\ }\bibfield  {title} {\bibinfo {title} {{Geometry and physics of wrinkling}},\ }\href {https://doi.org/10.1103/PhysRevLett.90.074302} {\bibfield  {journal} {\bibinfo  {journal} {Phys. Rev. Lett.}\ }\textbf {\bibinfo {volume} {90}},\ \bibinfo {eid} {074302} (\bibinfo {year} {2003})}\BibitemShut {NoStop}%
\bibitem [{\citenamefont {{Vella}}\ \emph {et~al.}(2006)\citenamefont {{Vella}}, \citenamefont {{Metcalfe}},\ and\ \citenamefont {{Whittaker}}}]{Vella2006}%
  \BibitemOpen
  \bibfield  {author} {\bibinfo {author} {\bibfnamefont {D.}~\bibnamefont {{Vella}}}, \bibinfo {author} {\bibfnamefont {P.~D.}\ \bibnamefont {{Metcalfe}}},\ and\ \bibinfo {author} {\bibfnamefont {R.~J.}\ \bibnamefont {{Whittaker}}},\ }\bibfield  {title} {\bibinfo {title} {{Equilibrium conditions for the floating of multiple interfacial objects}},\ }\href {https://doi.org/10.1017/S0022112005008013} {\bibfield  {journal} {\bibinfo  {journal} {J. Fluid Mech.}\ }\textbf {\bibinfo {volume} {549}},\ \bibinfo {pages} {215} (\bibinfo {year} {2006})}\BibitemShut {NoStop}%
\bibitem [{\citenamefont {Vella}\ \emph {et~al.}(2006)\citenamefont {Vella}, \citenamefont {Lee},\ and\ \citenamefont {Kim}}]{vella2006load}%
  \BibitemOpen
  \bibfield  {author} {\bibinfo {author} {\bibfnamefont {D.}~\bibnamefont {Vella}}, \bibinfo {author} {\bibfnamefont {D.-G.}\ \bibnamefont {Lee}},\ and\ \bibinfo {author} {\bibfnamefont {H.-Y.}\ \bibnamefont {Kim}},\ }\bibfield  {title} {\bibinfo {title} {The load supported by small floating objects},\ }\href@noop {} {\bibfield  {journal} {\bibinfo  {journal} {Langmuir}\ }\textbf {\bibinfo {volume} {22}},\ \bibinfo {pages} {5979} (\bibinfo {year} {2006})}\BibitemShut {NoStop}%
\bibitem [{\citenamefont {{Qu\'{e}r\'{e}}}(2008)}]{Quere2008}%
  \BibitemOpen
  \bibfield  {author} {\bibinfo {author} {\bibfnamefont {D.}~\bibnamefont {{Qu\'{e}r\'{e}}}},\ }\bibfield  {title} {\bibinfo {title} {{Wetting and roughness}},\ }\href@noop {} {\bibfield  {journal} {\bibinfo  {journal} {Annu. Rev. Mater. Res.}\ }\textbf {\bibinfo {volume} {38}},\ \bibinfo {pages} {71} (\bibinfo {year} {2008})}\BibitemShut {NoStop}%
\bibitem [{\citenamefont {Snoeijer}\ and\ \citenamefont {Andreotti}(2013)}]{Snoeijer2013}%
  \BibitemOpen
  \bibfield  {author} {\bibinfo {author} {\bibfnamefont {J.~H.}\ \bibnamefont {Snoeijer}}\ and\ \bibinfo {author} {\bibfnamefont {B.}~\bibnamefont {Andreotti}},\ }\bibfield  {title} {\bibinfo {title} {Moving contact lines: scales, regimes, and dynamical transitions},\ }\href@noop {} {\bibfield  {journal} {\bibinfo  {journal} {Annu. Rev. Fluid Mech.}\ }\textbf {\bibinfo {volume} {45}},\ \bibinfo {pages} {269} (\bibinfo {year} {2013})}\BibitemShut {NoStop}%
\bibitem [{\citenamefont {Joanny}\ and\ \citenamefont {Robbins}(1990)}]{Robbins1990}%
  \BibitemOpen
  \bibfield  {author} {\bibinfo {author} {\bibfnamefont {J.~F.}\ \bibnamefont {Joanny}}\ and\ \bibinfo {author} {\bibfnamefont {M.~O.}\ \bibnamefont {Robbins}},\ }\bibfield  {title} {\bibinfo {title} {Motion of a contact line on a heterogeneous surface},\ }\href@noop {} {\bibfield  {journal} {\bibinfo  {journal} {J. Chem. Phys.}\ }\textbf {\bibinfo {volume} {92}},\ \bibinfo {pages} {3206} (\bibinfo {year} {1990})}\BibitemShut {NoStop}%
\bibitem [{\citenamefont {Ren}\ and\ \citenamefont {Weinan}(2011)}]{Weinan2011}%
  \BibitemOpen
  \bibfield  {author} {\bibinfo {author} {\bibfnamefont {W.}~\bibnamefont {Ren}}\ and\ \bibinfo {author} {\bibfnamefont {E.}~\bibnamefont {Weinan}},\ }\bibfield  {title} {\bibinfo {title} {Contact line dynamics on heterogeneous surfaces},\ }\href@noop {} {\bibfield  {journal} {\bibinfo  {journal} {Phys. Fluids}\ }\textbf {\bibinfo {volume} {23}},\ \bibinfo {pages} {072103} (\bibinfo {year} {2011})}\BibitemShut {NoStop}%
\bibitem [{\citenamefont {Kim}\ \emph {et~al.}(2017)\citenamefont {Kim}, \citenamefont {Fezzaa}, \citenamefont {An}, \citenamefont {Sun},\ and\ \citenamefont {Jung}}]{Jung2017}%
  \BibitemOpen
  \bibfield  {author} {\bibinfo {author} {\bibfnamefont {S.~J.}\ \bibnamefont {Kim}}, \bibinfo {author} {\bibfnamefont {K.}~\bibnamefont {Fezzaa}}, \bibinfo {author} {\bibfnamefont {J.}~\bibnamefont {An}}, \bibinfo {author} {\bibfnamefont {T.}~\bibnamefont {Sun}},\ and\ \bibinfo {author} {\bibfnamefont {S.}~\bibnamefont {Jung}},\ }\bibfield  {title} {\bibinfo {title} {Capillary spreading of contact line over a sinking sphere},\ }\href@noop {} {\bibfield  {journal} {\bibinfo  {journal} {Appl. Phys. Lett.}\ }\textbf {\bibinfo {volume} {111}},\ \bibinfo {pages} {134102} (\bibinfo {year} {2017})}\BibitemShut {NoStop}%
\bibitem [{\citenamefont {Buryachenko}(2007)}]{Buryachenko2007}%
  \BibitemOpen
  \bibfield  {author} {\bibinfo {author} {\bibfnamefont {V.~A.}\ \bibnamefont {Buryachenko}},\ }\href@noop {} {\emph {\bibinfo {title} {Micromechanics of Heterogeneous Materials}}}\ (\bibinfo  {publisher} {Springer},\ \bibinfo {year} {2007})\BibitemShut {NoStop}%
\bibitem [{\citenamefont {Razavi}\ \emph {et~al.}(2015)\citenamefont {Razavi}, \citenamefont {Cao}, \citenamefont {Lin}, \citenamefont {Lee}, \citenamefont {Tu},\ and\ \citenamefont {Kretzschmar}}]{Razavi2015a}%
  \BibitemOpen
  \bibfield  {author} {\bibinfo {author} {\bibfnamefont {S.}~\bibnamefont {Razavi}}, \bibinfo {author} {\bibfnamefont {K.~D.}\ \bibnamefont {Cao}}, \bibinfo {author} {\bibfnamefont {B.}~\bibnamefont {Lin}}, \bibinfo {author} {\bibfnamefont {K.~Y.~C.}\ \bibnamefont {Lee}}, \bibinfo {author} {\bibfnamefont {R.~S.}\ \bibnamefont {Tu}},\ and\ \bibinfo {author} {\bibfnamefont {I.}~\bibnamefont {Kretzschmar}},\ }\bibfield  {title} {\bibinfo {title} {Collapse of particle-laden interfaces under compression: buckling vs particle expulsion},\ }\href@noop {} {\bibfield  {journal} {\bibinfo  {journal} {Langmuir}\ }\textbf {\bibinfo {volume} {31}},\ \bibinfo {pages} {7764} (\bibinfo {year} {2015})}\BibitemShut {NoStop}%
\bibitem [{\citenamefont {{Abkarian}}\ \emph {et~al.}(2007)\citenamefont {{Abkarian}}, \citenamefont {{Subramaniam}}, \citenamefont {{Kim}}, \citenamefont {{Larsen}}, \citenamefont {{Yang}},\ and\ \citenamefont {{Stone}}}]{Abkarian2007}%
  \BibitemOpen
  \bibfield  {author} {\bibinfo {author} {\bibfnamefont {M.}~\bibnamefont {{Abkarian}}}, \bibinfo {author} {\bibfnamefont {A.~B.}\ \bibnamefont {{Subramaniam}}}, \bibinfo {author} {\bibfnamefont {S.-H.}\ \bibnamefont {{Kim}}}, \bibinfo {author} {\bibfnamefont {R.~J.}\ \bibnamefont {{Larsen}}}, \bibinfo {author} {\bibfnamefont {S.-M.}\ \bibnamefont {{Yang}}},\ and\ \bibinfo {author} {\bibfnamefont {H.~A.}\ \bibnamefont {{Stone}}},\ }\bibfield  {title} {\bibinfo {title} {{Dissolution arrest and stability of particle-covered bubbles}},\ }\href@noop {} {\bibfield  {journal} {\bibinfo  {journal} {Phys. Rev. Lett.}\ }\textbf {\bibinfo {volume} {99}},\ \bibinfo {pages} {188301} (\bibinfo {year} {2007})}\BibitemShut {NoStop}%
\bibitem [{\citenamefont {Pitois}\ \emph {et~al.}(2015)\citenamefont {Pitois}, \citenamefont {Buisson},\ and\ \citenamefont {Chateau}}]{Pitois2015}%
  \BibitemOpen
  \bibfield  {author} {\bibinfo {author} {\bibfnamefont {O.}~\bibnamefont {Pitois}}, \bibinfo {author} {\bibfnamefont {M.}~\bibnamefont {Buisson}},\ and\ \bibinfo {author} {\bibfnamefont {X.}~\bibnamefont {Chateau}},\ }\bibfield  {title} {\bibinfo {title} {On the collapse pressure of armored bubbles and drops},\ }\href@noop {} {\bibfield  {journal} {\bibinfo  {journal} {Eur. Phys. J. E}\ }\textbf {\bibinfo {volume} {38}},\ \bibinfo {pages} {48} (\bibinfo {year} {2015})}\BibitemShut {NoStop}%
\bibitem [{\citenamefont {Kumaki}(1988)}]{Kumaki1988}%
  \BibitemOpen
  \bibfield  {author} {\bibinfo {author} {\bibfnamefont {J.}~\bibnamefont {Kumaki}},\ }\bibfield  {title} {\bibinfo {title} {Monolayer of polystyrene monomolecular particles on a water surface studied by langmuir-type film balance and transmission electron microscopy},\ }\href@noop {} {\bibfield  {journal} {\bibinfo  {journal} {Macromolecules}\ }\textbf {\bibinfo {volume} {21}},\ \bibinfo {pages} {749} (\bibinfo {year} {1988})}\BibitemShut {NoStop}%
\bibitem [{\citenamefont {Aveyard}\ \emph {et~al.}(2000)\citenamefont {Aveyard}, \citenamefont {Clint}, \citenamefont {Nees},\ and\ \citenamefont {Quirke}}]{Aveyard2000a}%
  \BibitemOpen
  \bibfield  {author} {\bibinfo {author} {\bibfnamefont {R.}~\bibnamefont {Aveyard}}, \bibinfo {author} {\bibfnamefont {J.~H.}\ \bibnamefont {Clint}}, \bibinfo {author} {\bibfnamefont {D.}~\bibnamefont {Nees}},\ and\ \bibinfo {author} {\bibfnamefont {N.}~\bibnamefont {Quirke}},\ }\bibfield  {title} {\bibinfo {title} {Structure and collapse of particle monolayers under lateral pressure at the octane/aqueous surfactant solution interface},\ }\href@noop {} {\bibfield  {journal} {\bibinfo  {journal} {Langmuir}\ }\textbf {\bibinfo {volume} {16}},\ \bibinfo {pages} {8820} (\bibinfo {year} {2000})}\BibitemShut {NoStop}%
\bibitem [{\citenamefont {Tolnai}\ \emph {et~al.}(2001)\citenamefont {Tolnai}, \citenamefont {Csempesz}, \citenamefont {Kabai-Faix}, \citenamefont {Kalman}, \citenamefont {Keresztes}, \citenamefont {Kovacs}, \citenamefont {Ramsden},\ and\ \citenamefont {H\'{o}rv\"{o}lgyi}}]{Tolnai2001}%
  \BibitemOpen
  \bibfield  {author} {\bibinfo {author} {\bibfnamefont {G.}~\bibnamefont {Tolnai}}, \bibinfo {author} {\bibfnamefont {F.}~\bibnamefont {Csempesz}}, \bibinfo {author} {\bibfnamefont {M.}~\bibnamefont {Kabai-Faix}}, \bibinfo {author} {\bibfnamefont {E.}~\bibnamefont {Kalman}}, \bibinfo {author} {\bibfnamefont {Z.}~\bibnamefont {Keresztes}}, \bibinfo {author} {\bibfnamefont {A.~L.}\ \bibnamefont {Kovacs}}, \bibinfo {author} {\bibfnamefont {J.~J.}\ \bibnamefont {Ramsden}},\ and\ \bibinfo {author} {\bibfnamefont {Z.}~\bibnamefont {H\'{o}rv\"{o}lgyi}},\ }\bibfield  {title} {\bibinfo {title} {Preparation and characterization of surface-modified silica-nanoparticles},\ }\href@noop {} {\bibfield  {journal} {\bibinfo  {journal} {Langmuir}\ }\textbf {\bibinfo {volume} {17}},\ \bibinfo {pages} {2683} (\bibinfo {year} {2001})}\BibitemShut {NoStop}%
\bibitem [{\citenamefont {Beltramo}\ \emph {et~al.}(2017)\citenamefont {Beltramo}, \citenamefont {Gupta}, \citenamefont {Alicke}, \citenamefont {Gunes}, \citenamefont {Baroud},\ and\ \citenamefont {Vermant}}]{Vermant2017}%
  \BibitemOpen
  \bibfield  {author} {\bibinfo {author} {\bibfnamefont {P.~J.}\ \bibnamefont {Beltramo}}, \bibinfo {author} {\bibfnamefont {M.}~\bibnamefont {Gupta}}, \bibinfo {author} {\bibfnamefont {A.}~\bibnamefont {Alicke}}, \bibinfo {author} {\bibfnamefont {I.~L. D.~Z.}\ \bibnamefont {Gunes}}, \bibinfo {author} {\bibfnamefont {C.~N.}\ \bibnamefont {Baroud}},\ and\ \bibinfo {author} {\bibfnamefont {J.}~\bibnamefont {Vermant}},\ }\bibfield  {title} {\bibinfo {title} {Arresting dissolution by interfacial rheology design},\ }\href@noop {} {\bibfield  {journal} {\bibinfo  {journal} {Proc. Natl. Acad. Sci. U.S.A.}\ }\textbf {\bibinfo {volume} {114}},\ \bibinfo {pages} {10373} (\bibinfo {year} {2017})}\BibitemShut {NoStop}%
\bibitem [{\citenamefont {{Jambon-Puillet}}\ \emph {et~al.}(2016)\citenamefont {{Jambon-Puillet}}, \citenamefont {{Vella}},\ and\ \citenamefont {{Proti{\`e}re}}}]{JambonPuillet2016}%
  \BibitemOpen
  \bibfield  {author} {\bibinfo {author} {\bibfnamefont {E.}~\bibnamefont {{Jambon-Puillet}}}, \bibinfo {author} {\bibfnamefont {D.}~\bibnamefont {{Vella}}},\ and\ \bibinfo {author} {\bibfnamefont {S.}~\bibnamefont {{Proti{\`e}re}}},\ }\bibfield  {title} {\bibinfo {title} {{The compression of a heavy floating elastic film}},\ }\href {https://doi.org/10.1039/C6SM00945J} {\bibfield  {journal} {\bibinfo  {journal} {Soft Matter}\ }\textbf {\bibinfo {volume} {12}},\ \bibinfo {pages} {9289} (\bibinfo {year} {2016})},\ \Eprint {https://arxiv.org/abs/1609.03366} {arXiv:1609.03366 [cond-mat.soft]} \BibitemShut {NoStop}%
\end{thebibliography}%

\end{document}